\documentclass{aa} 

\newcommand{\ltsim}{\protect\raisebox{-0.5ex}{$\:\stackrel{\textstyle <}
        {\sim}\:$}}

\usepackage{graphicx}
\usepackage[authoryear]{natbib}
\bibpunct{(}{)}{;}{a}{}{,} 

\begin{document}

\title{The young active star SAO\,51891 (V383~Lac)\thanks{Based on 
 observations collected at Calar Alto Astronomical Observatory (Spain) and Catania
Astrophysical Observatory (Italy).}}
   
\author{Katia Biazzo \inst{1,6,7} \and Antonio Frasca \inst{1} \and Ettore Marilli
\inst{1} \and Elvira Covino \inst{2} \and 
Juan M. Alcal\`a \inst{2} \and ${\rm \ddot{O}}$mur \c{C}akirli \inst{3} \and Alexis
Klutsch \inst{4} \and Michael R. Meyer \inst{5}}
\offprints{Katia Biazzo}
\mail{kbiazzo@arcetri.inaf.it}

\institute{INAF - Catania Astrophysical Observatory, via S. Sofia 78, I--95123 Catania, Italy
\and INAF - Capodimonte Astronomical Observatory, via Moiariello 16, I--80131 Napoli, Italy
\and Ege University, Science Faculty, Astronomy and Space Sciences Department, T--35100 Bornova, Izmir, Turkey
\and Strasbourg Astronomical Observatory, Universit\'e de Strasbourg \& CNRS (UMR 7550), 11 rue de 
l'Universit\'e, F--67000 Strasbourg, France \and Steward Observatory, The University of Arizona, 933 North Cherry Avenue, Tucson, AZ 85721-0065
\and INAF - Arcetri Astrophysical Observatory, Largo E. Fermi 5, I--50125 Firenze, Italy
\and ESO - European Southern Observatory, Karl-Schwarzschild-Str. 3, 85748 Garching bei M\"unchen, Germany
}

\date{Received / accepted }

\abstract
{}
{The aim of this work is to investigate the surface inhomogeneities of a young, late-type star, SAO\,51891, 
at different atmospheric levels, from the photosphere to the upper chromosphere, analyzing contemporaneous 
optical high-resolution spectra and broad-band photometry.} 
{The full spectral range of FOCES@CAHA ($R\,\simeq\,$40\,000) is used to perform the spectral classification and to 
determine the rotational and radial velocities. The lithium abundance is measured to obtain an age estimate. 
The $BVRIJHK_{\rm s}$ photometric bands are used to construct the spectral energy distribution (SED). The variations in 
the observed $BV$ fluxes and effective temperature are used to infer the presence of photospheric spots and observe their 
behavior over time. The chromospheric activity is studied applying the spectral subtraction technique to H$\alpha$, 
\ion{Ca}{ii} H\,\&\,K, H$\epsilon$, and \ion{Ca}{ii} IRT lines. }
{We find SAO\,51891 to be a young K0--1V star with a lithium abundance close to the Pleiades upper envelope, 
confirming its youth ($\sim 100\,$Myr), which is also inferred from its kinematical membership to the Local Association. 
No infrared excess is detected from analysis of its SED, limiting the amount of remaining circumstellar dust. 
We detect a rotational modulation of the luminosity, effective temperature, \ion{Ca}{ii} H\,\&\,K, H$\epsilon$, and 
\ion{Ca}{ii} IRT total fluxes. A simple spot model with two main active regions, about 240\,K cooler than the surrounding 
photosphere, fits very well the observed light and temperature curves. The small-amplitude radial velocity variations 
are also well reproduced by our spot model. The anti-correlation of light curves and chromospheric diagnostics indicates
chromospheric plages spatially associated with the spots. The largest modulation amplitude is observed for the 
H$\epsilon$ flux suggesting that this line is very sensitive to the presence of chromospheric plages.}
{SAO\,51891 is a young active star, lacking significant amounts of circumstellar dust or any evidence for low mass companions, 
which displays the typical phenomena produced by magnetic activity. The spots turn out to be larger and warmer than those in less
active main-sequence stars. If some debris material is still present around the star, it will only be detectable by future far-infrared 
and sub-mm observations (e.g., Herschel or ALMA). The RV variation produced by the starspots has an amplitude comparable with those 
induced by Jupiter-mass planets orbiting close to the host star. SAO\,51891 is another good example of an active star in which the detection of 
planets may be hampered by the high activity level.
}

   
\titlerunning{The young active star SAO\,51891}
\authorrunning{K. Biazzo et al.}
\maketitle

\section{Introduction}

\label{sec:Intro}

In the solar neighborhood there are several stellar kinematic groups (SKGs) including stars with a common space motion. 
The origin of these SKGs, like the Local Association (LA), or Pleiades moving group, could be the evaporation of an 
open cluster, the remnants of a star-forming region, or a superposition of several star formation bursts 
at different epochs in the adjacent cells of the velocity field (\citealt{Montes01a}). Therefore, the study of 
late-type stars in these young SKGs is important to better understand the recent star formation history of the 
solar neighborhood. 

Moreover, these young solar-mass stars just arrived on the Zero-Age Main Sequence (ZAMS), or on their way 
to it, are in a very important phase of their evolution. Indeed, as they approach the ZAMS, they spin up while getting 
free from their circumstellar disks, which eventually ``condense" giving rise to planetary systems. At the same time, 
the stars experience angular momentum loss resulting from a magnetized stellar wind. In these evolutionary phases, it 
is therefore very important to define the physical conditions that affect the subsequent evolution of the star and any 
possible planets. In addition to basic stellar parameters ($T_{\rm eff}$, $\log g$, [Fe/H]), it is necessary to know the 
rotation rate and the level and behavior of the magnetic activity. 

It is well known that the lower temperature in sunspot umbrae and penumbrae as well as in starspots is 
due to the blocking effect of emerging magnetic flux tubes over the convective energy transport 
in the sub-photospheric layers. The structure of a magnetic flux tube breaking into the stellar atmosphere 
is traced by the configuration of active regions (ARs) at different levels, like photospheric spots and chromospheric plages. 
Therefore, an accurate determination of spot temperatures and sizes such as that derived from the monitoring of line-depth ratios
(LDRs) along a complete rotation period (\citealt{Cata02}, \citealt{Frasca05}) and the associated variability of 
hydrogen, helium, and calcium lines (\citealt{Fra00,Frasca08}; \citealt{Biazzo06,Biazzo07b}) 
can provide crucial information on the magnetic field topology in young stars to be compared with the results for 
older main sequence (MS) and giant stars.

In the present work we apply this analysis to the young star \object{SAO\,51891} making use of contemporaneous 
$BV$ photometry and spectroscopic observations. The chromospheric activity was studied by means of the 
\ion{Ca}{ii} H\,\&\,K ($\lambda=$3968.49 \AA, 3933.68 \AA), H$\epsilon$ ($\lambda=$ 3970.074 \AA), H$\alpha$ ($\lambda=$6562.849 \AA), 
\ion{Ca}{ii} infrared triplet (IRT; 
$\lambda=$8498.06 \AA, 8542.14 \AA, 8662.17 \AA) lines. Our main aim is the determination of the spot temperature and filling factor and to
infer the distribution of chromospheric ARs of SAO\,51891. The general scope is the comparison of the results
obtained for very young stars, such as SAO\,51891, with those found for MS and RS~CVn stars (see, e.g.,
\citealt{Frasca05,Frasca08}; \citealt{Biazzo07b}).

Our work is organized as follows. In Section\,\ref{sec:sao51891} the main stellar parameters of SAO\,51891 are briefly
discussed. In Section\,\ref{sec:Obs} we give the details of the observations and data reduction. The spectral
classification, the rotational and radial velocities, the space motion, the spectral energy distribution, the position on the
HR diagram, the lithium abundance, and the metallicity are reported in Sections \ref{sec:spec_type}--\ref{sec:metall}.
The diagnostics of photospheric and chromospheric activity are analyzed in Sections \ref{sec:phot_act} and
\ref{sec:chrom_act}, while the spot model is presented in Section \ref{sec:spot_plage_model}. Section \ref{sec:concl}
contains the conclusions. 

\section{The case of SAO\,51891 (V383 Lac, BD+48 3686, 1RXS J222007.2+493014)}
\label{sec:sao51891}
SAO\,51891 was identified as the optical counterpart of the ROSAT extreme ultraviolet source RE\,J2220+493 \citep{Pounds91,Pye95}. 
Recent spectroscopic and photometric studies (\citealt{Mulliss94,Jeffries95,Henry95,Osten98,Xing2007}) concluded that it is a
single active K1V star slightly younger than the Pleiades and report $v\sin i$ values ranging from 14 to 20
km\,s$^{-1}$. In particular, \cite{Henry95} found the highest peaks in the periodogram of their
light-curve at P=2.42$\pm$0.01 days and 1.70$\pm$0.01 days, with the former stronger than the latter (both in $B$ and
$V$ bands). They found an amplitude of 0\fm08 with significant cycle-to-cycle variations of the magnitude
at minimum. These authors proposed also a minimum radius of $0.77\,R_\odot$ and an inclination
$i$ close to 90${\degr}$. This star does not show any relevant radial velocity variation as the mean value of
$-20.19$ km\,s$^{-1}$ obtained by \cite{Montes01a} appears consistent with the range of values (from
$-$19.4 to $-$22.1 km\,s$^{-1}$) reported in the literature. This suggests that SAO\,51891 is very likely a single star
(Table~\ref{tab:literature_param}). \cite{Mulliss94} found the H$\alpha$, \ion{Ca}{ii} H and \ion{Ca}{ii} IRT lines
filled-in with emission. \cite{Montes01a} reported notable emission in the \ion{Ca}{ii} H\,\&\,K and
H$\epsilon$ lines, excess chromospheric emission in the hydrogen Balmer lines and emission core reversal in the \ion{Ca}{ii}
IRT lines. On one of their observing nights, they detected an increase in the excess emission, with broad
emission wings in the H$\alpha$ residual profile. They explained this variation as probably due to a
small-scale flare or to the transit of an active region. They also determined a lithium equivalent width
($EW_{\rm \ion{Li}{i}}$) of 257 m\AA, similar to the values of 250 and 277 m\AA\ given by \cite{Mulliss94} and
\cite{Jeffries95}, respectively (Table~\ref{tab:literature_param}). This large $EW_{\rm \ion{Li}{i}}$ value, close
to the upper envelope of the Pleiades, indicates a young age. From its kinematics, SAO\,51891 has been considered
by \citet{Montes01a} as a member of the LA.

Infrared observations place limits on the amount of circumstellar dust present (see Section~\ref{sec:sed}). Of course 
these limits only probe dust grain populations with large emitting areas and do not place constraints on the presence 
of very low mass companions (or planets). \cite{Metchev2006} has conducted an adaptive optics (AOs) survey for faint 
companions in the vicinity of the star. Although several candidates were found, no co-moving companions were identified, 
placing limits on brown dwarf companions ($> 50\%$ complete) more massive than approximately the deuterium burning limit 
at $0.013\,M_{\odot}$ beyond approximately 20 AU. SAO~51891 was also a target of the near-infrared AO 
search for giant planets named ``Gemini Deep Planet Search'' 
(\citealt{Lafreniere2007}). Null results from their survey rule out potential planetary mass companions with masses greater
than $5\,M_{\rm J}$ beyond 25 AU with $58\%$ probability.

\begin{table}
\caption{Physical parameters of SAO\,51891 from the literature.}
\centering
 \begin{tabular}{lll}
  \hline\hline
  \noalign{\smallskip}
  Parameter                  & Value & Reference\\
  \noalign{\smallskip}
  \hline
  \noalign{\smallskip}
  $f_{\rm X}$ (cts\,s$^{-1}$)          & 1.03$\pm$0.05  & \cite{Voges1999}\\
  Spectral Type                   & K0V/IV & \cite{Jeffries95}\\
   "                              & K1V    & \cite{Henry95}\\
  $\pi$ (mas)                     & 42.4$\pm$7.2  & \cite{Hog2000} \\
   "                              & 36.3$\pm$5.0  & \cite{Montes01b} \\
   "                              & 20            & \cite{Carpenter2005} \\
  $\mu_{\alpha}$ (mas\,yr$^{-1}$) & 93.4$\pm$2.5  & \cite{Hog2000}\\
  $\mu_{\delta}$ (mas\,yr$^{-1}$) &  5.0$\pm$2.5  & \cite{Hog2000}\\
  $v\sin i$ (km\,s$^{-1}$)        & 15  	  & \cite{Jeffries95}\\
  "                               & $16\pm1$   & \cite{Henry95}\\
  "                               & 19.8       & \cite{Fekel97}\\
  "                               & 20         & \cite{Osten98}\\
  $P_{\rm rot}$ (days)            & 2.42$\pm$0.01     & \cite{Henry95}\\
  "                               & 2.410$\pm$0.003   & \cite{Xing2007}\\
  $R$ ($R_{\odot}$)               & $\ge$0.77 & \cite{Henry95}\\
  $EW_{\rm \ion{Li}{i}}$ (m\AA)   & 250                & \cite{Mulliss94}\\
  "                               & 277 		  & \cite{Jeffries95}\\
  "                               & 257 		  & \cite{Montes01a}\\
  $V_{\rm rad}$ (km\,s$^{-1}$)    & $-$19.4 & \cite{Jeffries95}\\
  "                               & $-19.8\pm0.2$   & \cite{Henry95}\\
  "                               & $-22.1\pm0.6$   & \cite{Osten98}\\
  "                               & $-20.19\pm0.12$ & \cite{Montes01a}\\
  $U_{\odot}~ ($km\,s$^{-1}$)     & $7.06\pm$1.43   & \cite{Montes01a} \\
  $V_{\odot}~ ($km\,s$^{-1}$)     & $-22.19\pm$0.34 & \cite{Montes01a} \\
  $W_{\odot}~ ($km\,s$^{-1}$)     & $-3.90\pm$0.86  & \cite{Montes01a} \\
  $T_{\rm eff}$ (K)               & 5200  & \cite{Osten98}\\
  \noalign{\smallskip}
  \hline
\end{tabular}
\label{tab:literature_param}
\end{table}

\section{Observations and data reduction}

\label{sec:Obs}

\subsection{Spectroscopy}
The spectroscopic observations were conducted in 2006 at the 2.2-m Cassegrain telescope 
of the German-Spanish Calar Alto Observatory (CAHA, Sierra de Los Filabres, Spain) during
four nights from 13 to 16 August. The Fiber Optics Cassegrain {\it \'Echelle} Spectrograph (FOCES; 
\citealt{Pfeiffer1998}) was used with the 2048$\times$2048 CCD detector Site\#1d 
(pixel size = 24\,$\mu$m, read-out-noise=5 e$^{-}$ r.m.s., conversion factor = 2.3\,e$^{-}$/ADU) 
allowing us to achieve, with an exposure time of 35 minutes, a signal-to-noise ratio ($S/N$) in 
the range 70--100, depending on airmass and sky conditions. 
The wavelength range covered by the 88 {\it \'echelle} orders was 3720--8850 \AA. 
We observed SAO\,51891 at least twice per night obtaining in total 10 spectra of the
object.

We chose the Site\#1d CCD for its higher quantum efficiency in the wavelength range of 
our interest and the 200\,$\mu$m fiber-mode in order to avoid substantial light losses 
and maximize the $S/N$.

The spectral resolution, determined from the full width at half maximum (FWHM) of the emission lines of the 
Th-Ar calibration lamp, was in the range 0.15--0.22\,\AA\ from the blue to the red, yielding a 
resolving power $R=\lambda/\Delta\lambda\,\simeq\,$40\,000. 

The data reduction was performed by using the {\sc echelle} task of the IRAF\footnote{IRAF is distributed by the 
National Optical Astronomy Observatory, which is operated by the Association of the Universities for Research in 
Astronomy, inc. (AURA) under cooperative agreement with the National Science Foundation.} package following the 
standard steps of background subtraction, division by a flat-field spectrum given by a halogen lamp, wavelength 
calibration using the emission lines of the Th-Ar lamp, and normalization to the continuum through a 
polynomial fit. 

We removed the telluric water vapor lines near H$\alpha$ using the spectra of 
\object{Altair} (A7\,V, $v\sin i=245$~km\,s$^{-1}$) acquired during the observing run, and by 
applying an interactive procedure, which allowed the intensity of the template lines 
to vary (leaving the line ratios unchanged) until a satisfactory match 
with each of the observed spectra was achieved \citep[see][]{Fra00}.

\subsection{Photometry}
The photometric observations were performed in the $B$ and $V$ Johnson filters with the 91-cm 
Cassegrain telescope at the {\it M. G. Fracastoro} station ({\it Serra La Nave}, Mt. Etna, Italy) of the 
{\it Osservatorio Astrofisico di Catania} (OACt). The observations were made with a photon-counting 
refrigerated photometer equipped with an EMI 9893QA/350 photomultiplier, cooled to $-15 ^\circ$C. 
The dark current noise of the detector, operated at this temperature, is about $1$ count/sec.
A 21$\arcsec$ diaphragm was used.

SAO\,51891 was observed in 2006 from 14 to 21 August for a total of 8 nights, using \object{BD+48 3666} ($V=8\fm308$, 
$B-V=0.484$) and \object{HD~12028} ($V=8\fm520$, $B-V=0.050$) as comparison ($C$) and check ($Ck$) stars, 
respectively. Additionally, local photometric standard stars, as well as stars of some Landolt's selected 
areas \citep{Lan92} were also observed during the run in order to determine the zero points and the transformation 
coefficients to the Johnson standard system, respectively. 
The differential magnitudes, in the sense of variable minus comparison ($V-C$), were corrected for atmospheric 
extinction using the seasonal average coefficients for the {\it Serra La Nave} Observatory. 

A measurement on each object consisted of two integration cycles with a 3-sec exposure time in the $B$ and 
$V$ filters, and a typical observing sequence, {\em sky--Ck--C--V--C--V}, was repeated many times
per each night. The data were reduced by means of the photometric data reduction package {\sc phot} designed for 
the photoelectric photometry of the OACt \citep{LoPr93}. The photometric errors, estimated from 
measurements of standard stars with a brightness comparable to the program stars, are typically 
$\sigma_V \approx 0\fm010$ and $\sigma_{B-V} \approx 0\fm014$.

\section{Data analysis and results}

\begin{table}
\caption{Physical parameters of SAO\,51891 derived in this work.}
\centering
 \begin{tabular}{ll}
  \hline\hline
  \noalign{\smallskip}
  Parameter                  & Value \\
  \noalign{\smallskip}
  \hline
  \noalign{\smallskip}
  Spectral Type                  & K0-1$\pm$1\,V \\
  $(B-V)_0$                      & 0.785$\pm$0.014 \\
  $V_0$                          & $8\fm485\pm0\fm010$ \\
  $A_{\rm V}$                    & 0.017\\
  $P_{\rm rot}$                  & 2.62$\pm$0.45 days \\
  $L$                            & $0.215\pm0.075\,L_\odot$ \\
  $[$Fe/H$]$                     & 0.28$\pm$0.05\\
  $\log g^{\rm SED}$             & 4.0$\pm$0.1 \\
  $\log g^{\rm SPEC}$            & 4.3$\pm$0.1 \\
  $T_{\rm eff}^{\rm SED}$        & 5420$\pm$150 K \\
  $T_{\rm eff}^{\rm SPEC}$       & 5350$\pm$100 K \\
  $T_{\rm eff}^{\rm LDR}$        & 5249$\pm$70 K \\
  $v\sin i$                      & 19.0$\pm$0.5 km\,s$^{-1}$ \\
  $R\sin i$                      & $0.8\pm0.1\,R_\odot$  \\
  $i$                            & 75${\degr}\pm$15${\degr}$ \\
  $M$                            & $0.8\pm0.2\,M_\odot$ \\
  $EW_{\rm \ion{Li}{i}}$         & 260$\pm$10 m\AA \\
  $V_{\rm rad}$                  & $-$19.67$\pm$0.24 km\,s$^{-1}$ \\
  $U_{\odot}$                    & $+5.68\pm$1.46 km\,s$^{-1}$ \\
  $V_{\odot}$                    & $-21.32\pm$0.46 km\,s$^{-1}$ \\
  $W_{\odot}$                    & $-3.05\pm$1.01 km\,s$^{-1}$ \\
  Age                            & $\ltsim$100 Myr \\
  \noalign{\smallskip}
  \hline
\end{tabular}
\begin{flushleft}
\end{flushleft}
\label{tab:target_param}
\end{table}

\subsection{Spectral classification and rotational velocity}
\label{sec:spec_type}
To derive the spectral type and the $v\sin i$ of the target, we used {\sc rotfit}, a code written by 
\citet{Frasca03} in the IDL\footnote{IDL (Interactive Data Language) is a trademark of ITT Visual Information 
Solutions (ITT VIS).} environment and successfully applied by us for the spectral classification of single-lined 
active binaries in the RasTyc sample of stellar X-ray sources (\citealt{Frasca2006}). The code simultaneously 
finds the spectral type and the $v\sin i$ of the target searching for, into a library of standard star spectra, 
the spectrum which gives the best match (minimum of the residuals) with the target one, after the rotational 
broadening by convolution with a rotational profile of increasing $v\sin i$ at steps of 0.5 km\,s$^{-1}$. 
During our observing run we acquired spectra of only nine standard stars with different spectral type. 
Thus, we decided to use a library of ELODIE\footnote{This {\it \'echelle} spectrograph of the 1.93-m telescope 
of the {\it Observatoire de Haute-Provence} is now decommissioned.} 
Archive spectra (\citealt{Prugniel01}) of 87 standard stars well 
distributed in effective temperature, spectral type, and gravity, and in a suitable range of metallicities. 
For the $v\sin i$ determination, we chose as a ``non-rotating'' template 
the FOCES spectrum of 54\,Psc (K0\,V, $v\sin i=2.2\pm0.8$ km\,s$^{-1}$; \citealt{Fekel97}) acquired by us during the
observing run with the same instrumental set-up. This allowed us to avoid small differences between FOCES and ELODIE
spectra, although the latter have nearly the same resolution ($R\simeq\,42\,000$) as the former. Anyway, the average
value of $v\sin i$ found with ELODIE templates is exactly the same as that found with the FOCES spectrum of 54\,Psc
($v\sin i=19$\,km\,s$^{-1}$).

\begin{figure*}
\centering
\includegraphics[width=14cm]{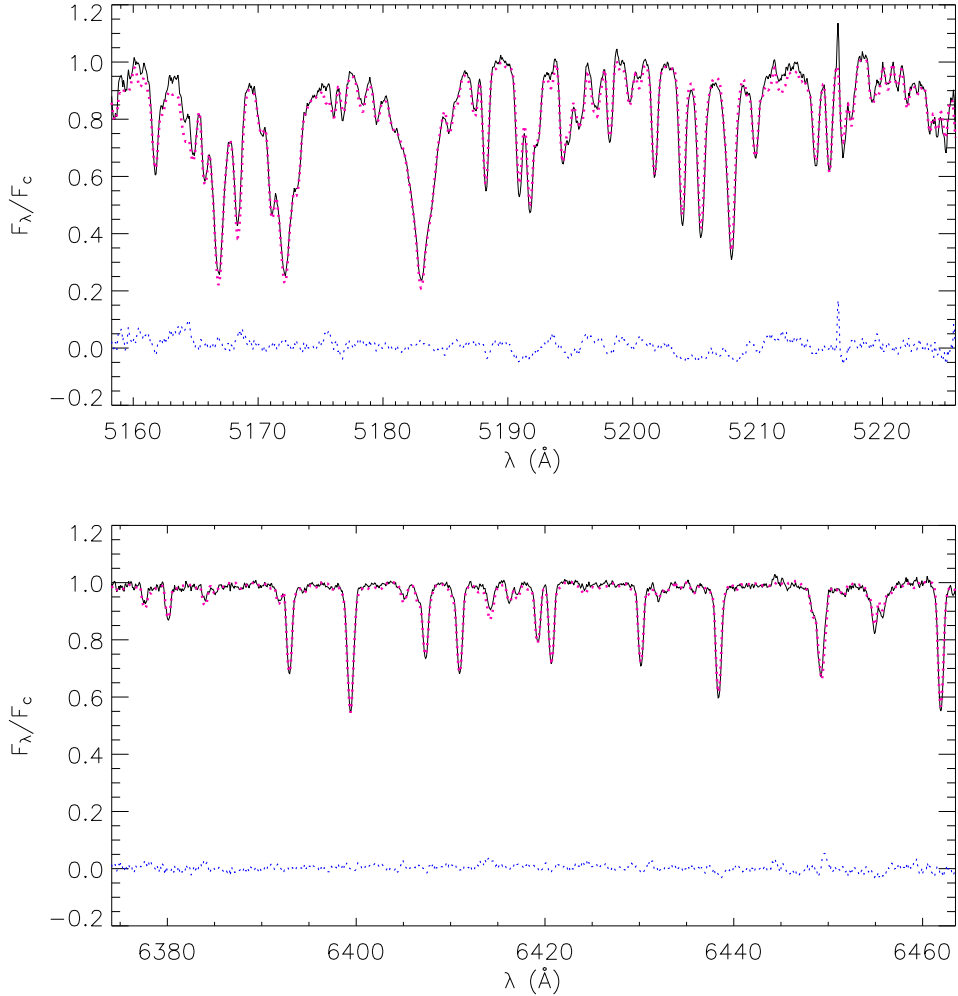}
\caption{Observed FOCES spectrum of SAO\,51891 in the \ion{Mg}{i} triplet (upper panel) and 6400 {\AA} (lower panel) 
spectral regions together with the ELODIE standard spectrum broadened at 19 km\,s$^{-1}$ overplotted with a 
dashed line. In each box the difference (observed $-$ synthetic) is also displayed with a dotted line.} 
\label{fig:spect_cl1}
\end{figure*}

Examples of application of the {\sc rotfit} code to two different spectral regions are shown in 
Fig.~\ref{fig:spect_cl1}, where the good agreement between observed and standard spectra is evident. 
The spectral type we find is K0--1$\pm$1\,V and we obtain for $v\sin i$ a best-fit value
of 19.0$\pm$0.5 km\,s$^{-1}$ (Table~\ref{tab:target_param}), near to the larger values from the literature. 

\subsection{Spectral Energy Distribution}
\label{sec:sed}
\begin{figure}
\centering
\includegraphics[width=9cm]{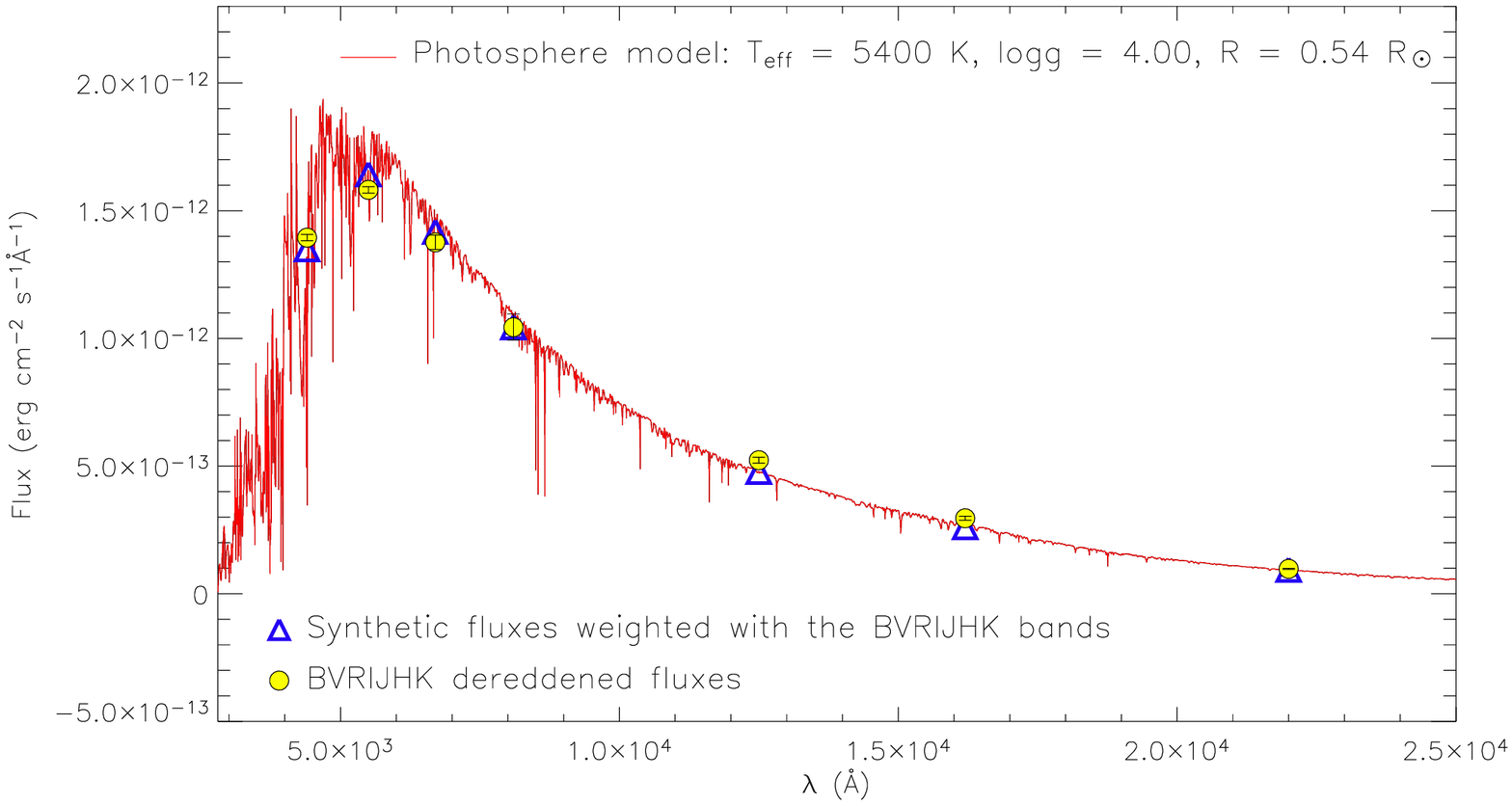}
\includegraphics[width=9cm]{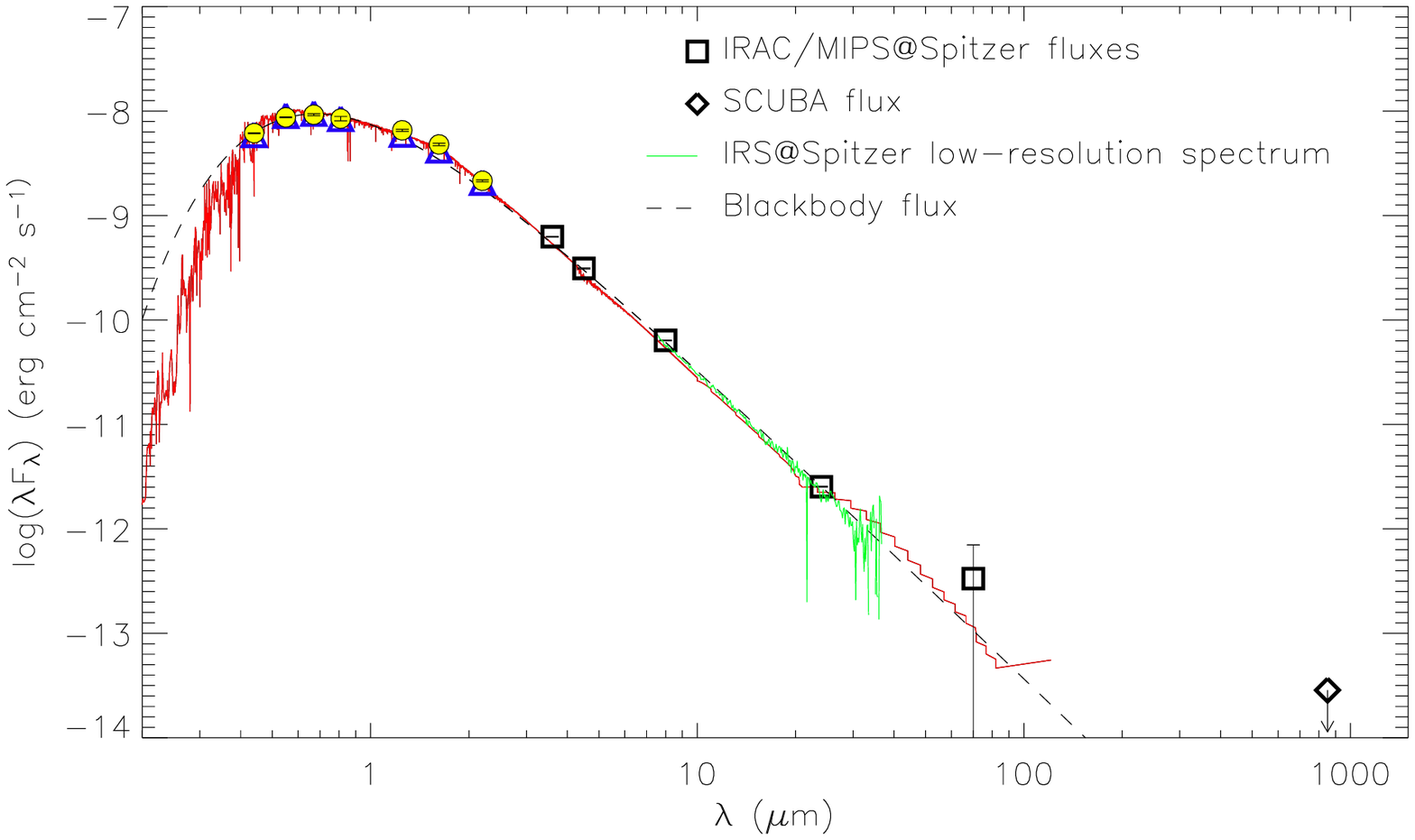}
\caption{{\it Top panel:} SED of SAO\,51891 as deduced from $BVRIJHK_{\rm s}$ observed magnitudes 
(open circles) compared to the synthetic fluxes (triangles) by integrating in the proper passbands a synthetic 
NextGen (full line) model for $T_{\rm eff}=5400$ K and $\log g^{\rm SED}=4.0$. $B$ and $V$ data are referred to 
our observations, while $R$, $I$, and $JHK_{\rm s}$ data were taken from NOMAD (\citealt{Zacha2004}), TASS 
(\citealt{Droege2006}), and 2MASS (\citealt{Cutri2003}) catalogues, respectively. {\it Bottom panel:} Same 
figure shown in log-log scale, where the IRAC/MIPS@{\it Spitzer} (at 3.6, 4.5, 8, 24, 70 $\mu$m; \citealt{Carpenter2008}) 
and SCUBA/JCMT (at 850 $\mu$m; \citealt{Najita2005}) fluxes are also displayed. The dashed 
line represents the result obtained considering a black body at $T_{\rm eff}=5400$ K, 
the continuous line in the 7.6--37 $\mu$m range is instead the IRS@{\it Spitzer} 
spectrum (\citealt{Carpenter2008}). 
}
\label{fig:sed}
\end{figure}

We defined the spectral energy distribution (SED) in the optical/infrared domain with the
$BVRIJHK_{\rm s}$ observed magnitudes after a standard de-reddening (\citealt{Cox00}). 
The $B$ and $V$ magnitudes are those obtained by us at OACt near the maximum brightness (minimum spot 
visibility; cf. Section \ref{sec:light_curve}), while the $R$, $I$, and 
$JHK_{\rm s}$ magnitudes come from the Naval Observatory Merged Astrometric Dataset (NOMAD; \citealt{Zacha2004}), 
The Amateur Sky Survey (TASS; \citealt{Droege2006}), and Two Micron All Sky Survey (2MASS; \citealt{Cutri2003}) 
catalogues, respectively. These observed fluxes were compared with those obtained from the low-resolution 
synthetic NextGen spectra (\citealt{Haus99}) integrated over the pass-bands of the filters. 
The model with $T_{\rm eff}=5400$ K, $\log g=4.0$, and solar metallicity gives the best fit to the observed 
SED. The radius derived from the fit of the SED and adopting the Tycho distance is $R=0.54\pm0.09\,R_{\odot}$, where the 
error is essentially due to the distance uncertainty. This value is in very close agreement with that coming from 
the \cite{Barnes1976} relation and the Tycho distance ($R=0.55\pm0.19\,R_{\odot}$). 
However, this value of $R$ is at odds with the mass-radius relation for MS stars. 
Indeed, according to the tabulation of \citet{Cox00}, a K0\,V star should have a $T_{\rm eff}=5150$\,K, 
in agreement with our findings, but a radius of 0.85\,$R_{\sun}$ that is much larger than the value 
implied by the Tycho distance. Moreover, if we combine the available determinations of projected rotational velocity, 
and the rotation period of the star, we derive a lower limit for the stellar radius, $R\sin i$, 
of $0.9\,R_{\odot}$ (using our determination of v$\sin i = 19$\,km\,s$^{-1}$), or $0.7\,R_{\odot}$ 
(if we adopt the smallest value, v$\sin i = 15$\,km\,s$^{-1}$ from literature). 
Nevertheless, if we use the distance of 27.5\,pc reported by \citet{Montes01b}, the stellar radius turns out 
to be $R=0.64\,R_{\odot}$, thus reducing the discrepancy. The distance of 50\,pc reported 
by \citet{Carpenter2005} would instead increase the radius to $R\approx1.1\,R_{\odot}$. 
Hence, we see good reasons to suspect that the Tycho distance is underestimated and a better 
distance determination, like that expected from the future mission GAIA of ESA, is needed to settle this point.
In any event, the photospheric temperature is only dependent on the shape of the SED and the distance has a very 
small effect on it through the interstellar extinction, which is, in any case, very low. 
Interpolating the NextGen fluxes at 10\,K steps, the best solution in temperature, coming from 
the $\chi^2$ minimization, is 5420\,K (Table~\ref{tab:target_param}). In Fig.~\ref{fig:sed}, 
we show the observed SED in both linear and log scales, the flux computed from the model, and the 
best-fit synthetic spectrum.   For comparison, Carpenter et al. (2008) report an 
effective temperature of 5120 $\pm 130$ K with a reddening $A_V$ = 0, assuming solar metallicity. 

SAO\,51891 was also observed within the {\it Spitzer} Legacy Program ``Formation and Evolution of Planetary Systems" 
(FEPS; \citealt{Meyer2006}), which is a comprehensive study of the evolution of gas and dust surrounding sun-like 
stars from the pre-main sequence (PMS) phase to the age of the Sun. Data from 3.6--70\,$\mu$m plotted 
in Fig.~\ref{fig:sed} are from this survey. The ratio of 24\,$\mu$m emission to 8\,$\mu$m emission reported by \cite{Carpenter2008} 
is consistent with that expected from the stellar photosphere (\citealt{Meyer2008}). Comparing the upper limit on any potential 
excess emission to the models of \cite{Hines2006} suggests a limit on the warm dust mass of $< 10^{-5}\,M_\oplus$ in 
1-10\,$\mu$m grains. SAO 51891 is not detected by {\it Spitzer} at 70\,$\mu$m at a significant level (\citealt{Carpenter2008}). 
Comparing the $3\sigma$ upper limit with the 70\,$\mu$m flux to the emission expected from the star, we can estimate 
$F(70$-disk$)/F(70$-star$) < 16$. Comparing this limit to debris disk models of early K stars (e.g., HD~31392) from 
\cite{Hillenbrand2008} suggests a limit to the cool (50 K) dust mass of $<10^{^-4}\,M_\oplus$ in 10\,$\mu$m grains. An 
upper limit to the flux at 850\,$\mu$m, from a survey by \citet{Najita2005} using the Submillimiter Common-User Bolometric 
Array (SCUBA) at the James Clerk Maxwell Telescope (JCMT) on Mauna Kea (Hawaii), is also displayed. More sensitive upper 
limits on any cold dust will require additional far-IR observations with the future  mission Herschel of ESA 
or mm-wave interferometers such as ALMA.

\subsection{Position on the HR diagram}
\label{sec:HR_diagram}

\begin{figure}
\centering
\includegraphics[width=9.2cm]{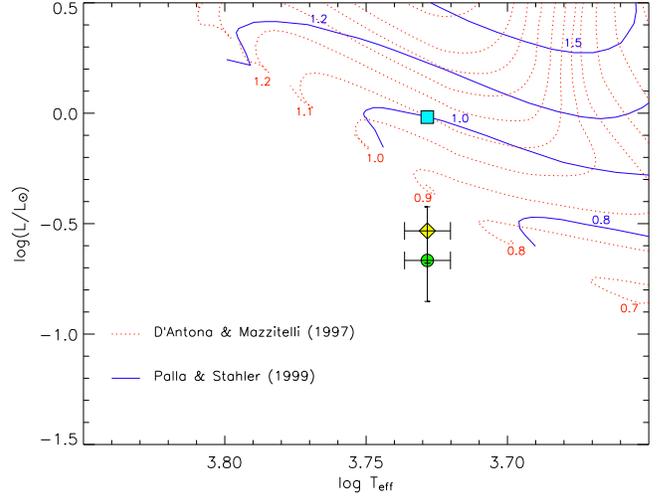}
\caption{HR diagram of SAO\,51891. A dot, a diamond, and  a square are used for the values of luminosity derived
from the Tycho, \citet{Montes01b}, and \citet{Carpenter2005} distances, respectively. The evolutionary tracks of 
\citet{palla1999} are shown by continuous lines. Dotted lines are instead used for the tracks of 
\citet{dant_mazz1997}.} 
\label{fig:hr_diagram}
\end{figure}

To check the evolutionary status of SAO\,51891, we placed it onto the Hertzsprung-Russel (HR) 
diagram (Fig.~\ref{fig:hr_diagram}) and compared its position with the PMS evolutionary 
tracks calculated by \citet{palla1999} and \citet{dant_mazz1997}. 
The mean effective temperature of SAO\,51891 was derived considering three determinations: 
i) $5420\pm150$\,K obtained interpolating the NextGen flux values at 10\,K steps (cf. Section \ref{sec:sed}); 
ii) the ``spectroscopic'' value of $5350\pm100$\,K obtained imposing the condition that the \ion{Fe}{i} 
abundance does not depend on the excitation potentials of the iron lines (cf. Section \ref{sec:metall}); 
iii) the ``LDR-method'' value of $5249\pm70$\,K corresponding to the maximum of the temperature-curve, 
representative of the unspotted (or less spotted) photosphere (the error was evaluated as the quadratic 
sum of the mean error of $\approx 10$\,K on the individual $T_{\rm eff}$ values and the rms error of the 
LDR--$T_{\rm eff}$ calibration, cf. Section~\ref{sec:teff_ldr}). At the end, a mean value of $5350\pm100$\,K 
was adopted. The stellar luminosity was derived according to the relation $L=4\pi R^2\sigma T_{\rm eff}^2$, 
where the radius obtained from the SED and assuming the Tycho distance is $R=0.54\pm0.09\,R_{\odot}$.
This yields a value of $L=0.215\pm0.075$\,L$_{\odot}$ which places SAO\,51891 beneath the ZAMS (Fig.\,\ref{fig:hr_diagram}).
Therefore, we calculated the luminosity also using the distance values reported by 
\citet{Montes01b} and \citet{Carpenter2005} finding $L=0.293\pm0.084\,L_{\odot}$ and $L=0.960\,L_{\odot}$, respectively.
These values, displayed with different symbols in the HR diagram (Fig.\,\ref{fig:hr_diagram}), appear to
be more reliable. We thus conclude that the Tycho distance is most probably underestimated.

With the latter luminosity values, the position of SAO\,51891 in the HR diagram is consistent with a ZAMS 
or a post-T~Tauri star, in agreement with the lithium content and kinematics (cf. Sections 
\ref{sec:radial_velocity} and \ref{sec:lithium_age}). We derive a mass of about $0.9\pm0.1\,M_{\odot}$.

\subsection{Radial Velocity and space motion}
\label{sec:radial_velocity}
The heliocentric radial velocity (RV) measurements were obtained by means of the cross-correlation 
technique (e.g., \citealt{Simkin1974}, \citealt{Gunn1996}) using the IRAF task {\sc fxcor}. 
A FOCES spectrum of the RV standard star $\alpha$\,Ari ($RV_{\rm helio}=-14.2$ km\,s$^{-1}$; \citealt{Evans1967}) was 
used as a template. In order to take advantage of the wide spectral coverage offered by FOCES, we cross-correlated 
each spectral order of the SAO\,51891 spectra with the template, excluding only the orders with low $S/N$ ratio 
(e.g., the 80th and 81st, which furthermore include the \ion{Ca}{ii} H and K lines) or contaminated by broad 
and/or chromospheric lines (e.g., H$\alpha$, \ion{Na}{ii}, and \ion{Ca}{ii} IRT) or by prominent telluric features (e.g., the O$_2$ series at
$\lambda\simeq6275$\,\AA). Ultimately, about 60 orders were considered for the cross-correlation function (CCF). In order to better
evaluate the centroids of the CCF peaks, we adopted Gaussian fits. The standard errors of the RV values in each order
were computed using the {\sc fxcor} task according to the fitted peak height and the antisymmetric noise as
described by \cite{TonryDavis1979}. For each spectrum, the RV values from individual orders were averaged with weights
$w_i = 1/\sigma_i^2$ (where $\sigma_i$ is the RV error for the $i$-th order). The resulting RVs are listed in
Table~\ref{tab:eqw_lines}. The average RV value over the entire observing run is $RV=-19.67\pm0.24$ km\,s$^{-1}$
(Table~\ref{tab:target_param}), which is close to the recent determination obtained by \cite{Montes01a}. 
As shown in Fig.~\ref{fig:palle_teff_V_err_sao} and in Table~\ref{tab:eqw_lines}, RV differences up to 
0.51 km\,s$^{-1}$ were measured in our spectra. Since this value of RV amplitude is about 
three times the average error on individual measurements, we consider it significant. These 
variations are correlated with the rotational phase and are ascribable to line-asymmetry changes caused 
by starspots rather than to an unseen planetary companion (see Section~\ref{sec:spot_plage_model}).

\begin{figure*}
\includegraphics[width=17cm]{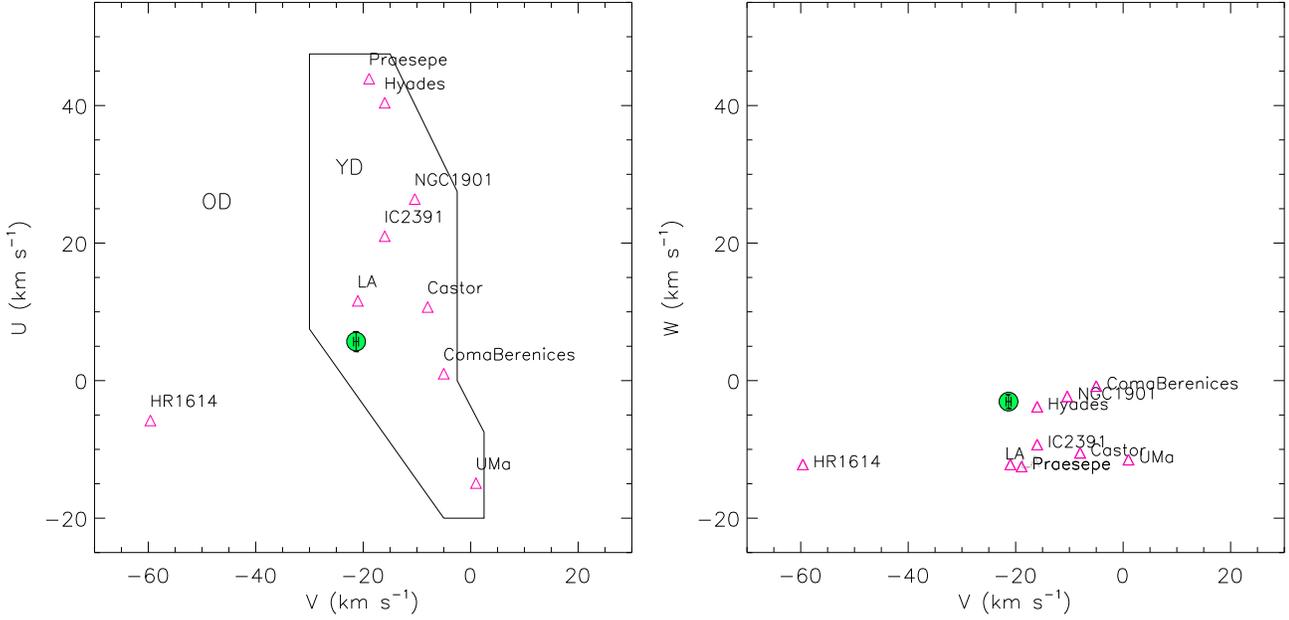}
\caption{Space velocities of SAO\,51891 in the ($U,V$) and ($W,V$) planes. The continuous line represents the 
boundary separating young-disk (YD) and old-disk (OD) stars according to \citet{Eggen1996}. The average 
velocity components of some moving groups (\citealt{Eggen1996}, and references therein) are also marked
by triangles.} 
\label{fig:space_motion_sao}
\end{figure*}

We used our RV determination combined with the parallax and proper motion from the Tycho-2 catalogue 
(Table~\ref{tab:literature_param}) to derive the Galactic space-velocity components ($U_{\odot}$, 
$V_{\odot}$, $W_{\odot}$) in a left-handed coordinate system (positive in the direction of the 
Galactic anti-center, the Galactic rotation, and the North Galactic Pole). We considered the general outline 
of \cite{JohSod1987} with the FK4 coordinates at epoch=1950. The uncertainty was obtained using the full 
covariance matrix taking into account the error contributions of $V_{\rm rad}$, $\mu_{\alpha}$, $\mu_{\delta}$, 
and $\pi$. The values derived are listed in Table~\ref{tab:target_param} and the Boettlinger Diagrams in the 
($U,V$) and ($W,V$) planes are plotted in Fig.~\ref{fig:space_motion_sao}, where the position of some young SKG 
is also displayed. Our ($U,V,W$) values are very close to those found by \cite{Montes01a}, who derived 
$U=7.06\pm1.43$ km\,s$^{-1}$, $V=-22.19\pm0.34$ km\,s$^{-1}$, and $W=-3.90\pm0.86$ km\,s$^{-1}$ as Galactic 
space-velocity components. We remark that the other values of parallax listed in 
Table\,\ref{tab:literature_param} do not change significantly the values of the space motion components. 
We also applied a kinematic method developed by \cite{Klutsch2008} to determine 
the membership probability to five young SKGs, namely, the IC2391 supercluster, the Pleiades moving group or 
Local Association, the Castor moving group, the Ursa Major group or Sirius supercluster, and the Hyades 
supercluster. We find a space velocity fully consistent with the young-disk population and a high membership probability 
(64\%) to the LA (20--150 Myr). Moreover, SAO\,51891 is placed inside the subgroup B1 of age $20\pm10$ Myr associated 
with the Pleiades cluster identified by \cite{Asiain1999}. 

\subsection{Lithium abundance and age}
\label{sec:lithium_age}
The lithium abundance was evaluated from the equivalent width of the $\lambda6707.8$ line, $EW_{\rm \ion{Li}{i}}$.
The latter was measured using the IRAF task {\sc splot}, and its error was computed by multiplying the integration 
range and the mean error per spectral point evaluated at the continuum on the two sides of the line.
The lithium abundance was derived using the curve of growth (COG) method of \cite{Soderblom1993}. The contribution 
due to the \ion{Fe}{i} line at $\lambda$6707.4 was subtracted using the empirical relationship given by the same 
authors: $\Delta EW_{\rm Li} ({\rm m\AA})=20(B-V)_0-3$. Lithium abundances were then corrected for NLTE 
effects using the prescriptions of \cite{Carlsson1994}. In the case of our target, the LTE lithium abundance 
turns out to be $\log N_{\rm Li}^{\rm LTE}\approx$3.18 dex, while the NLTE effects contribute 
almost halving it ($\log N_{\rm Li}^{\rm NLTE}\approx$2.95 dex). The lithium abundance we derive is more 
than a hundred times the solar value ($\log N_{\rm Li,\odot}^{\rm LTE}=1.1\pm0.1$ dex; \citealt{Loddersetal2009}).

In the diagram $T_{\rm eff}$--$EW_{\rm \ion{Li}{i}}$, SAO\,51891 lies above the Pleiades upper envelope, indicating 
an age around one hundred Myr. This age corresponds to a post-T Tauri (PTT) or ZAMS evolutionary stage.
Since our target is a single star, its photospheric and chromospheric activity (cf. Section~\ref{sec:phot_act}) as 
well as its strong coronal X-ray emission (\citealt{Voges1999}) should be essentially the effect 
of its young age (see the $f_{\rm X}$ value in Table~\ref{tab:literature_param}).

At this stage, most solar-mass stars are fast or ultra-fast rotators, as observed in the Pleiades and other young
open clusters (see, e.g., \citealt{Marilli97,Barnes2003}, and references therein), although a fraction of slow rotators is 
also observed. This is attributed to different evolutionary processes, such as a larger effectiveness or duration  
of the disk-locking mechanism while the star contracts towards the ZAMS. As such, SAO\,51891 appears to be somewhat
similar to the slow rotators in young open clusters.

\subsection{Metallicity}
\label{sec:metall}
The iron abundance calculations were performed in LTE conditions with an updated and improved version of the original 
code described in \cite{Spite1967}. \cite{Edvardsson1993} model atmospheres were used. Equivalent widths 
of spectral lines were measured using the {\sc splot} task in IRAF. The line list and corresponding atomic data 
were adopted from \cite{Pasquini2004}. The effective temperature was determined by imposing the condition that 
the \ion{Fe}{i} abundance does not depend on the excitation potentials of the lines. The microturbulence velocity 
$\xi$ was determined by imposing that the \ion{Fe}{i} abundance is independent of the line equivalent widths. The 
surface gravity $\log g$ was determined by imposing the \ion{Fe}{i}/\ion{Fe}{ii} ionization equilibrium. The initial 
value for the effective temperature was the one obtained by LDRs ($T_{\rm eff}^{\rm LDR}$; Section~\ref{sec:teff_ldr}). 
The initial value of $\log g$ was obtained taking into account the evolutionary tracks of \citet{palla1999}, using 
the $T_{\rm eff}^{\rm LDR}$, solar metallicity and $M_{\rm v}$ determined by our photometry and Tycho parallax 
measurements. Due to the good quality of the spectra and the many lines present in our wide spectral range, we 
could avoid to use lines in the flat portion of the curve of growth (\citealt{daSilva2006}). The initial 
microturbulence velocity was set to 1.0 km\,s$^{-1}$. We derived a metallicity $[$Fe/H$]$ of 0.28$\pm$0.05, a 
temperature $T_{\rm eff}^{\rm SPEC}=5350\pm100$ K, a surface gravity $\log g^{\rm SPEC}=4.3\pm0.1$ and a 
microturbulence $\xi=1.8$ km\,s$^{-1}$ (Table~\ref{tab:target_param}).

\subsection{Photospheric activity}
\label{sec:phot_act}

\subsubsection{Light-curve}
\label{sec:light_curve}
The photometric data acquired contemporaneously to the spectroscopic ones allowed us to obtain a light-curve 
showing a rotational modulation due to the presence of spots (Table~\ref{tab:phot_data}, 
Fig.~\ref{fig:palle_teff_V_err_sao}). The rotational phases were derived according to the following ephemeris:
\begin{equation}
HJD_{\phi=0} = 2\,453\,961.25+2\fd42\,\times\,E\,,
\label{Eq:ephem_lambda}
\end{equation}
{\noindent where the initial heliocentric Julian day corresponds to the first observing date (namely August 
13$^{\rm st}$, 2006) at 18:00 UT (just before our first observation) and the rotational period ($P_{\rm rot}$) 
of 2.42 days is taken from \cite{Henry95}. This period is close to the value of $2.410\pm0.003$ days recently 
found by \cite{Xing2007}. With a periodogram analysis and a {\sc clean} deconvolution algorithm (\citealt{Roberts}) 
we find a period of $2.62\pm0.45$ days, which is close the value found by \cite{Henry95} and \cite{Xing2007}, 
taking into account the errors. We decided to consider the period computed by \cite{Henry95}, because their 
estimation was based on the largest data-set (Table~\ref{tab:literature_param}).

The $V$ light-curve has a slightly asymmetric shape with a peaked top ($V_{\rm max}=8\fm495$) and a variation 
amplitude $\Delta V$=0\fm074. The $B$ light-curve has a shape similar to that obtained in the $V$ band, with 
$\Delta B=0\fm079$. The $B-V$ color shows a scattered modulation with an amplitude of only 0\fm025, 
which is at a low level of detectability ($\sim2\sigma$). To evaluate the significance of the 
$B-V$ variation, we performed a $\chi^2$ test by fitting a periodic function
(Fourier polynomial), following the scheme proposed by \cite{Biazzo06}. Notwithstanding the low amplitude,
the $B-V$ modulation turns out to be significant. However, it can hardly be used to constrain the spot
parameters and an additional information, such as that provided by the line-depth ratios, must be used
for this purpose.
}

\subsubsection{Effective temperature from line-depth ratios}
\label{sec:teff_ldr}
\begin{figure}
\centering
\includegraphics[width=11cm]{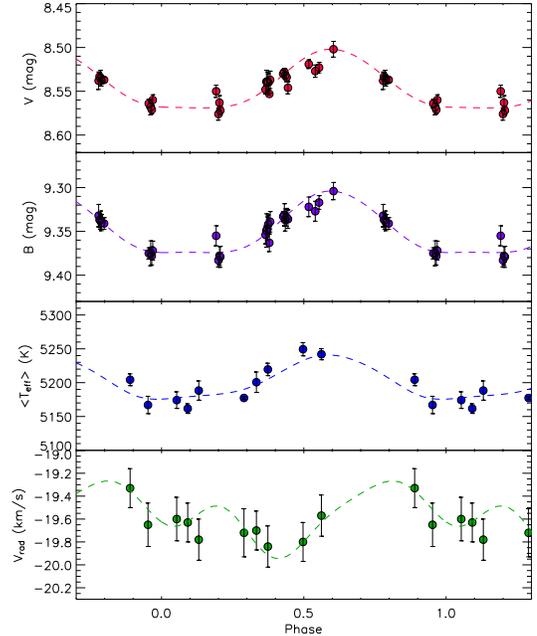}
\vspace{-2.3cm}
\caption{Observed (dots) and synthetic (dashed lines) $V$, $B$, $T_{\rm eff}$, $V_{\rm rad}$ curves of 
SAO\,51891 versus the rotational phase.} 
\label{fig:palle_teff_V_err_sao}
\end{figure}

It has been demonstrated that line-depth ratios (LDRs) are powerful tools for detecting temperature rotational modulations 
in stars with moderate magnetic activity (e.g., \citealt{Biazzo07b}) and with a high level of activity (e.g., 
\citealt{Cata02,Frasca05,Frasca08}). Such diagnostics allow us to detect temperature variations as small as 10\,K at 
the resolution of our spectra and with a signal-to-noise ratio higher than 100 (\citealt{Gray1991}). The precision of 
this method is improved by averaging the results from several line pairs. In particular, for SAO\,51891 we used 
fifteen line pairs in the optical spectral range 6200--6300 \AA\ and the LDR--$T_{\rm eff}$ calibrations developed 
by \cite{Biazzo07a} at $v\sin i=20$ km\,s$^{-1}$. Thus, we obtained the temperature-curve as the weighted 
average of all the temperature-curves from each LDR (Fig.~\ref{fig:palle_teff_V_err_sao}). 
The standard error of the weighted average was computed on the basis of the errors in each LDR-derived 
temperature. As shown in Fig.~\ref{fig:palle_teff_V_err_sao}, the average effective temperature is in phase 
with the light-curve, with only a possible shift of the maximum by less than 0\fp1. 
However, we have a gap in the decreasing part of the temperature-curve which casts doubts on the reality 
of this shift. The flat temperature minimum, with a dip around phase 0\fp0, coincides in phase with the minimum in the 
light-curve. This fair coherence confirms the hypothesis of cool spots as the primary cause of the observed variations. 
The temperature varies between 5162\,K and 5249\,K, with a full amplitude of about 90\,K, which is 
intermediate between the values of $\approx$\,40\,K found in moderately active stars (\citealt{Biazzo07b}) 
and $\approx$\,130\,K found in highly active stars (\citealt{Cata02,Frasca05,Frasca08}).

\subsection{Chromospheric activity}
\label{sec:chrom_act}
The wide wavelength range of FOCES spectra allowed us to study the chromosphere of SAO\,51891 by using several 
lines from the near UV to the NIR wavelengths (namely, \ion{Ca}{ii} H\,\&\,K, H$\alpha$, \ion{Ca}{ii} IRT), 
which carry information on different atmospheric levels, from the region of temperature minimum to the upper chromosphere. 
To derive the chromospheric losses, we used the ``spectral synthesis" technique, based on the comparison between the target 
spectrum and an observed spectrum of a non-active standard star (called ``reference spectrum"). The difference between the 
observed and the reference spectrum provides, as residual, the net chromospheric line emission, 
which can be integrated to find the total radiative losses in the line \citep[see, e.g.,][]{Herb85, Bar85, Fra94, Montes95}. 

\begin{figure}
\centering
\includegraphics[width=8.5cm]{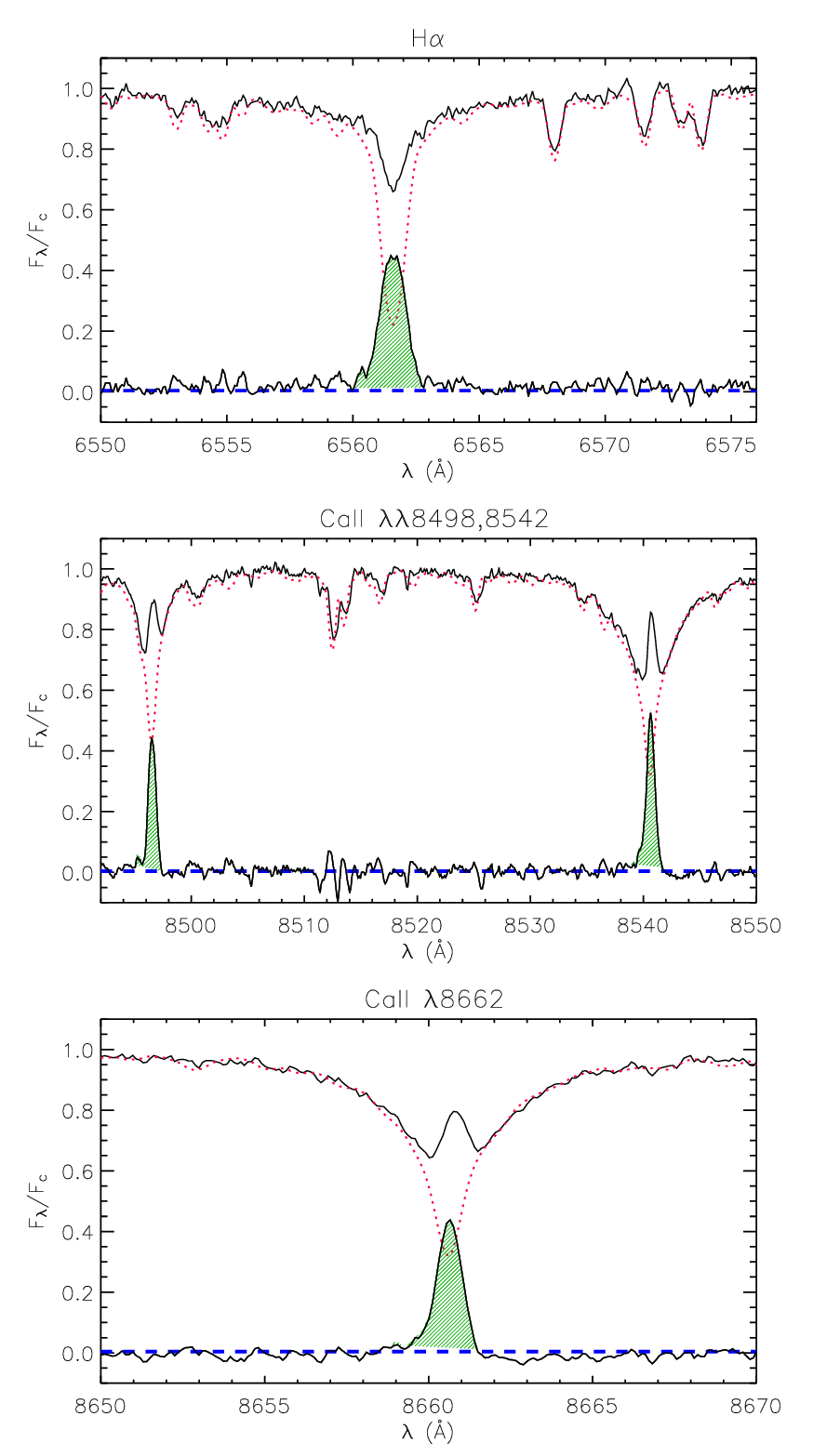}
\caption{{\it Top of each panel}: Examples of observed, continuum-normalized spectra of SAO\,51891 
(solid line) in the H$\alpha$ and \ion{Ca}{ii} IRT regions together with the non-active 
stellar template (dotted line). {\it Bottom of each panel}: The difference spectra of the two upper spectra.} 
\label{fig:SAO51891_spectra}
\end{figure}

\begin{figure*}
\centering
\includegraphics[width=14cm]{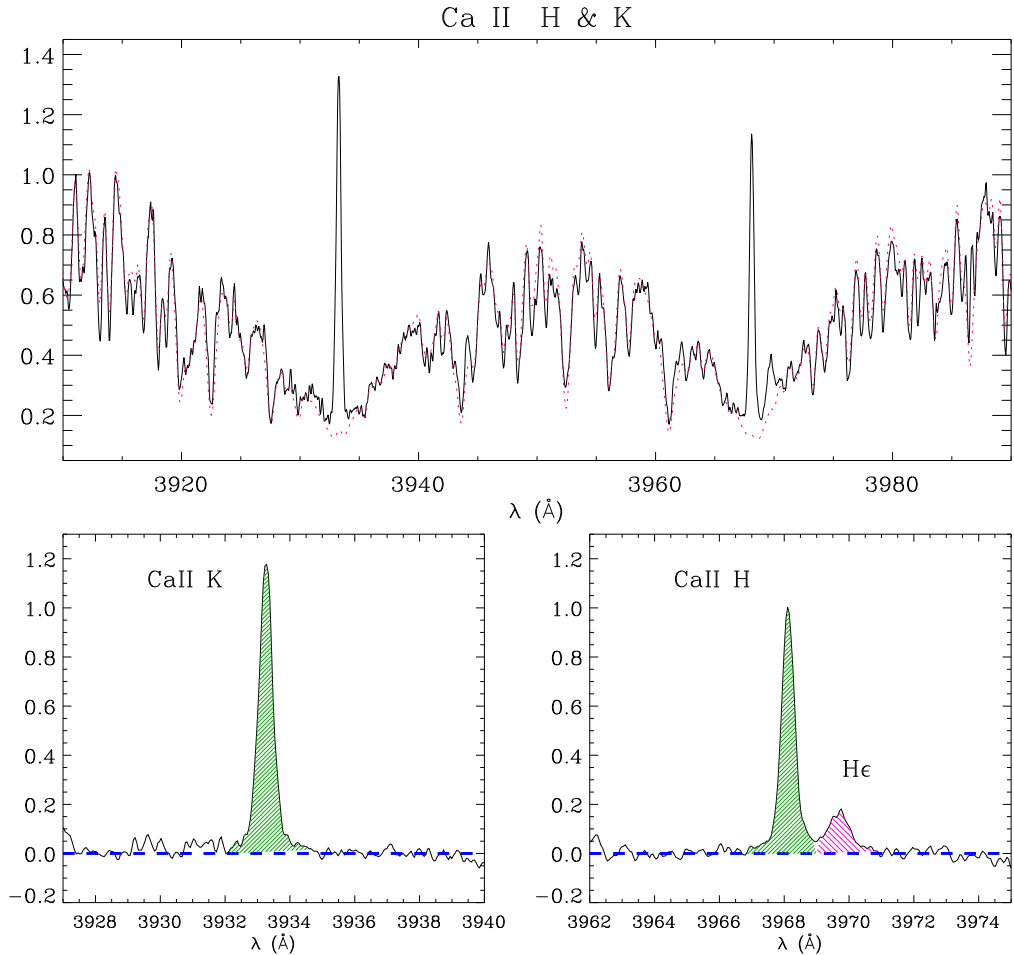}
\caption{{\it Top panel}: Example of an observed, continuum-normalized spectrum of SAO\,51891 (solid line) 
in the \ion{Ca}{ii} H\,\&\,K region together with the non-active stellar template (dotted line). 
{\it Bottom panels}: The difference spectrum in the \ion{Ca}{ii} K and H regions. H$\epsilon$ emission is 
clearly visible both in the raw and in the difference spectrum.} 
\label{fig:CaIIHK}
\end{figure*}

The non-active star used as a reference for the spectral subtraction is \object{54~Psc} (=\object{HD~3651}), a K0\,V star 
($B-V=0.85$) with a very low activity level, as indicated by the low value of the \ion{Ca}{ii} $S$ index 
(0.195; \citealt{Duncan1991}) and by the low flux in the \ion{Ca}{ii}\,K line measured on high-resolution
spectra (namely $\log \pi F_{\rm K}=4.98$; \citealt{Cata79}). This star was also observed during the same run as SAO\,51891. 
In Fig.~\ref{fig:SAO51891_spectra}, we show an example of a spectrum of SAO\,51891 in the H$\alpha$, 
and \ion{Ca}{ii} IRT regions, together with the standard-star spectrum rotationally broadened to $v\sin i=19$ km\,s$^{-1}$ 
which mimics the active star in absence of chromospheric activity. The \ion{Ca}{ii} H\,\&\,K region is displayed in 
Fig.~\ref{fig:CaIIHK}. The H$\alpha$ and \ion{Ca}{ii} IRT profiles are clearly filled-in by emission, with the calcium 
lines displaying an emission core. The \ion{Ca}{ii} H\,\&\,K cores exhibit strong emission features and H$\epsilon$ 
emission is also clearly visible (Fig.~\ref{fig:CaIIHK}). 

The residual equivalent width ($EW$) of the lines has been measured by integrating all the emission profile 
in the difference spectrum (see bottom of each panel of Fig.~\ref{fig:SAO51891_spectra} and bottom panels 
of Fig.~\ref{fig:CaIIHK}). The error on $EW$ ($\sigma_{EW}$) was evaluated by multiplying the integration range 
by the photometric error on each point. The latter was estimated by the standard deviation of the observed fluxes 
on the difference spectra in two spectral regions near the line. 

\subsubsection{H$\alpha$ line}
It has been widely shown, both from theoretical and observational points of view, that the H$\alpha$ line 
is one of the most useful and easily accessible indicators of chromospheric emission related to solar and 
stellar activity. Being its source function photoionization-dominated in the quiet chromospheres of the Sun and 
solar-like stars, this line contains valuable informations on the chromospheric structure and it is very sensitive 
to the strong non-thermal velocity. The H$\alpha$ absorption in the spectra of cool stars provides evidence for 
the existence in these stars of chromospheres with significant optical depth in that line. 
As a consequence, H$\alpha$ is useful for detecting chromospheric solar and stellar plages, thanks to their 
high contrast against the surrounding chromosphere.

The H$\alpha$ line of SAO\,51891 is always in absorption with a strong filling-in of the core that appears broader than
that observed in MS stars with mild activity like $\kappa_1$\,Cet \citep{Biazzo07b} or HD\,206860 \citep{Fra00}.
Moreover, the residual H$\alpha$ profile of SAO\,51891 is almost symmetric and does not show a significant variation 
(see Fig.~\ref{fig:SAO51891_spectra} and Appendix \ref{appendix:a}).

The values of the residual emission $EW_{\rm H\alpha}$ with their errors are listed in Table~\ref{tab:eqw_lines} 
and plotted as a function of the rotational phase in Fig.~\ref{fig:SAO51891_Halpha_He_CaII_Heps}. No rotational 
modulation of $EW_{\rm H\alpha}$ emerges over the data scatter. This result maybe is surprising, because rotational 
modulation of the H$\alpha$ emission has been frequently detected in several highly- and mildly-active stars
\citep[see, e.g.,][]{Fra00,Frasca08b, Alek02,Alek03, Biazzo06,Biazzo07b}. 
However, in some cases, no clear rotational modulation of chromospheric diagnostics has been detected despite 
the contemporaneous wave-like behaviour in the photometry (see, e.g., \citealt{Cata00}).
In such cases, the distribution of the active regions, as revealed by the H$\alpha$ emission, may be more homogeneously 
distributed in the base of the chromosphere than in the photosphere or in other atmospheric layers, as proposed, e.g., 
by \citet{Garcia-Alvarez2003}. In addition, the H$\alpha$ line may originate in other physical processes, like flares and prominences.

Since we observe a scatter in the H$\alpha$ EW of SAO~51891 larger than the typical data errors, we suppose that additional phenomena, such as microflaring, 
producing intrinsic variations of the emission can affect the line profile, hiding any low-amplitude rotational modulation. 
The rotational modulation of \ion{Ca}{ii} H\&K and H$\epsilon$ line fluxes (cf. \ref{sec:ca_hk} and \ref{sec:ca_irt})
supports this idea.

We also calculated the chromospheric radiative losses in the H$\alpha$ line, $F_{\rm H\alpha}$, following 
the guidelines by \citet{Fra94}, i.e. by multiplying the net equivalent width, $EW_{\rm H\alpha}$, by the continuum 
surface flux at $\lambda=6563$\,\AA. The latter was evaluated for all the stars of the spectrophotometric atlas 
of \citet{GunnStryker} with the angular diameters calculated from the \citet{Barnes1976} relation. The continuum 
flux of SAO\,51891 was found by interpolating the values for the \citet{GunnStryker} stars at $B-V=0.785$. 
The line flux $F_{\rm H\alpha}$ is reported in Table~\ref{tab:fluxes} together with those of other chromospheric 
diagnostics (see Sections \ref{sec:ca_hk} and \ref{sec:ca_irt}).

\begin{table*}[t]   
\caption{Radial velocities, effective temperatures, and parameters of the subtracted spectra.}
\label{tab:eqw_lines}
\scriptsize
\begin{center}
\begin{tabular}{ccccccccc}
\hline
\hline
  \noalign{\smallskip}
HJD & Phase & $V_{\rm rad}$& $T_{\rm eff}$ & $EW_{\rm H\alpha}$ & $EW_{\rm \ion{Ca}{ii}}^{\rm IRT}$ & 
$EW^{\rm H+K}_{\ion{Ca}{ii}}$ & $EW_{\rm H\epsilon}$ \\ 
(+2\,400\,000) & & (km\,s$^{-1}$)& (K) & (\AA) & (\AA) & (\AA) & (\AA) \\ 
  \noalign{\smallskip}
\hline
  \noalign{\smallskip}
53\,961.379 & 0.053 & $-19.60\pm$0.19 & 5174$\pm$12& 0.579$\pm$0.036& 1.424$\pm$0.055 &                 &      \\
53\,961.473 & 0.092 & $-19.63\pm$0.17 & 5162$\pm$7 & 0.633$\pm$0.028& 1.394$\pm$0.054 & 1.168$\pm$0.059 & 0.175$\pm$0.023\\
53\,961.566 & 0.131 & $-19.78\pm$0.18 & 5188$\pm$14& 0.615$\pm$0.038& 1.406$\pm$0.051 & 1.195$\pm$0.055 & 0.226$\pm$0.022\\
53\,962.453 & 0.497 & $-19.80\pm$0.17 & 5249$\pm$10& 0.625$\pm$0.030& 1.404$\pm$0.054 & 1.116$\pm$0.046 & 0.141$\pm$0.015\\
53\,962.609 & 0.562 & $-19.57\pm$0.18 & 5242$\pm$8 & 0.572$\pm$0.036& 1.409$\pm$0.059 & 1.036$\pm$0.068 & 0.116$\pm$0.019\\
53\,963.402 & 0.889 & $-19.33\pm$0.17 & 5204$\pm$9 & 0.603$\pm$0.023& 1.361$\pm$0.053 & 1.107$\pm$0.052 & 0.134$\pm$0.017\\
53\,963.555 & 0.952 & $-19.65\pm$0.19 & 5167$\pm$13& 0.656$\pm$0.036& 1.337$\pm$0.067 & 1.228$\pm$0.069 & 0.143$\pm$0.019\\
53\,964.371 & 0.290 & $-19.72\pm$0.21 & 5177$\pm$2 & 0.560$\pm$0.035& 1.431$\pm$0.071 & 1.289$\pm$0.119 & 0.205$\pm$0.042\\
53\,964.477 & 0.333 & $-19.70\pm$0.17 & 5201$\pm$13& 0.594$\pm$0.021& 1.419$\pm$0.060 & 1.209$\pm$0.060 & 0.144$\pm$0.015\\
53\,964.574 & 0.374 & $-19.84\pm$0.18 & 5220$\pm$9 & 0.673$\pm$0.028& 1.464$\pm$0.060 & 1.230$\pm$0.053 & 0.169$\pm$0.014\\
  \noalign{\smallskip}
\hline
\hline
\end{tabular}
\end{center}
\end{table*}
\normalsize

\begin{table}
\caption{Radiative chromospheric losses.}
\centering
 \begin{tabular}{lc}
  \hline\hline
  \noalign{\smallskip}
  Line                       & Flux \\			     
                             & (erg\,cm$^{-2}$\,s$^{-1}$) \\ 
  \noalign{\smallskip}
  \hline
  \noalign{\smallskip}
  H$\alpha$                  &  3.0$\times10^6$ \\
  \ion{Ca}{ii} H+K           &  4.7$\times10^6$ \\
  \ion{Ca}{ii} IRT           &  4.9$\times10^6$ \\
  \ion{Mg}{ii} h+k           &  4.9$\times10^6$ \\
  \noalign{\smallskip}
  \hline\hline
\end{tabular}
\begin{flushleft}
\end{flushleft}
\label{tab:fluxes}
\end{table}

\subsubsection{\ion{Ca}{ii} H\,\&\,K and H$\epsilon$ lines}
\label{sec:ca_hk}
The \ion{Ca}{ii} H\,\&\,K lines show very strong emission cores without any detectable self-absorption. 
They also appear very symmetric in all spectra. The H$\epsilon$ emission is clearly visible at the right side 
of the \ion{Ca}{ii} H line, both in the observed and in the residual spectra (Fig.~\ref{fig:CaIIHK}).

Since the absorption wings of each of the two lines span two {\it \'echelle} orders, it was necessary 
to merge them before to proceed with the spectral subtraction analysis. 
For the order merging and normalization we used the same procedure as in \citet{Fra00}.
The spectrum of the low-activity star \object{54~Psc}, broadened at $v\sin i$=19\,km\,s$^{-1}$, was used as a non-active
template, as explained in Section~\ref{sec:chrom_act}. Although a tiny emission/filling-in of the core of the K line is barely 
visible, it is negligible compared with the huge emission of SAO\,51891. Moreover, we preferred to use the same 
template for all the activity diagnostics.

We measured the equivalent widths by integrating the emission profiles in the subtracted spectra, as for H$\alpha$. 
For deblending the \ion{Ca}{ii} H and H$\epsilon$ lines, we performed fits of two Voigt functions by means of
the {\sc splot} task of IRAF. Both the H and K $EW$s display a fair rotational modulation, nearly anti-correlated 
with the light-curve. In Fig.~\ref{fig:SAO51891_Halpha_He_CaII_Heps}, we show the sum of the net equivalent widths, 
$EW^{\rm H+K}_{\ion{Ca}{ii}}$. A more pronounced modulation is observed in the net equivalent width of the H$\epsilon$ line.

\citet{Turova1994} observed a long-lasting H$\epsilon$ emission ($\approx 6$ days) above umbral regions for a sunspot group. 
This is a very rare phenomenon in the Sun, observed sometimes during flares. Based on the intensity ratio 
H$\epsilon$/\ion{Ca}{ii}\,H, she suggested that the sunspot group behaved like a dKe or dMe star, and concluded that 
the emission in the H$\epsilon$ line is related to an enhancement of collisional transitions owed to a very large 
temperature gradient in the chromosphere (\citealt{Fosbury1974}). \citet{Fosbury1974} stressed the fact that the 
emission in H$\epsilon$ could be due to a photoionization-recombination driven by the \ion{Ca}{ii}\,H line and background 
continuum radiation for stars with low chromospheric density, like in Arcturus, but the line is collision-dominated in 
dKe and dMe where the electron density is much higher. Moreover, when the electron density in the chromosphere is high, 
the \ion{Ca}{ii} emission cores are saturated and the relative intensity $I_{\rm H}/I_{\rm K}$ approaches unity. In 
these conditions, the H$\epsilon$ emission becomes more prominent compared to that of \ion{Ca}{ii} H. Fig.\,2 of 
\citet{Fosbury1974} displays the \ion{Ca}{ii} H\,\&\,K region of AD Leo, in which the H$\epsilon$ emission peak is 
more than one third of the \ion{Ca}{ii} H intensity ($\approx 0.35-0.40$). For SAO\,51891 we find values in the 
range $0.16-0.19$ with a possible (very scattered) modulation. Hence, if the chromosphere above an active region 
is strongly non-thermally heated and the density is high enough, H$\epsilon$ goes into emission and enhances much more 
than the \ion{Ca}{ii} lines do. This could explain the high amplitude of the H$\epsilon$ modulation and suggests 
this line as a sensitive marker of chromospheric surface features.

We also calculated the radiative chromospheric losses in the H\,\&\,K lines analogously as done for H$\alpha$.
In particular, we used two 10\,\AA-wide bands centered at 3910 and 4010~\AA, i.e. at the two sides of the H\,\&\,K 
lines to perform the flux calibration, following the prescriptions by \citet{Fra00}.

Finally, we evaluated the radiative losses in the h\,\&\,k lines of \ion{Mg}{ii} on a low-resolution UV 
spectrum, namely LWP26439LL.FITS, the only one available in the IUE\footnote{International Ultraviolet Explorer} 
Final Archive, and using as non-active template a resampled IUE spectrum of 54\,Psc. The radiative losses in 
the \ion{Ca}{ii} H\,\&\,K and \ion{Mg}{ii} h\,\&\,k, which are also listed in Table~\ref{tab:fluxes}, turn out 
to be nearly the same.

\subsubsection{\ion{Ca}{ii} IRT lines}
\label{sec:ca_irt}
These lines share the upper level of the H and K transitions and are useful for stellar chromospheric activity 
studies. Their extended wings probe a wide range of photospheric layers and are sensitive to the temperature 
distribution in the atmosphere of the star, while their cores are formed in the uppermost atmospheric layers 
and are sensitive to the physical conditions in the chromosphere. Some investigations based on 
empirical chromospheric models applied 
to active stars on a wide range of activity levels suggest that the contribution of the \ion{Ca}{ii} IRT lines can be 
up to twice larger than the contribution of the \ion{Ca}{ii} H\,\&\,K lines (\citealt{Dempsey1993}). 
As in the Sun, the calcium lines significantly contribute to the total chromospheric losses of the active stars, 
providing useful information on the energy balance of stellar chromospheres (\citealt{Busa07}, and references therein).

The lines of the \ion{Ca}{ii} infrared triplet present some advantages compared to the \ion{Ca}{ii} H\,\&\,K lines. 
First of all, the triplet lines lie in a spectral region with a well-defined continuum, making the normalization 
easier during the data reduction. Moreover, they are not significantly affected by telluric lines and are less 
affected by atmospheric extinction than the visible and ultraviolet lines. Finally, the high sensitivity of 
the new detectors to the near-IR makes these lines more easily observable than in the past. 
For this reason the calcium triplet has become a very versatile tool and the spectral region 8480--8740 \AA\ has been 
selected for the medium resolution spectrograph of the GAIA mission, as pointed out by \cite{Busa07}.

In SAO\,51891 the \ion{Ca}{ii} IRT lines are always filled-in with a central emission peak never reaching the 
local continuum. The profiles of the two lines $\lambda\lambda$ 8542.14,8662.17 are almost 
symmetric, while the $\lambda$8498.06 line displays an asymmetric profile 
(Figs.~\ref{fig:SAO51891_spettri2006_CaII8500}, \ref{fig:SAO51891_spettri2006_CaII8662}).

\begin{figure}
\centering
\includegraphics[width=9cm]{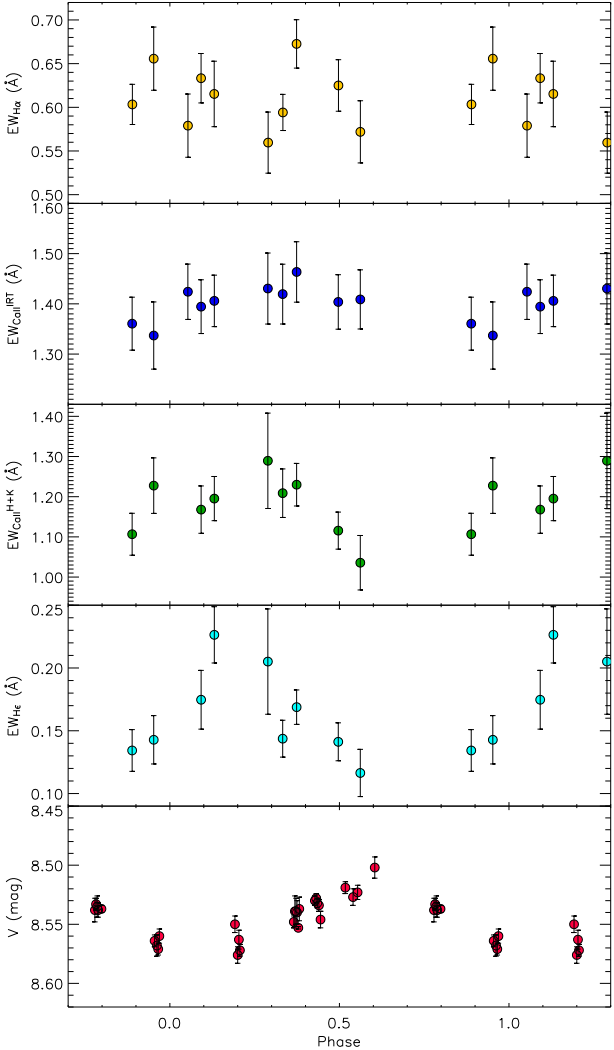}
\caption{{\it From top to bottom.} $EW_{\rm H\alpha}$, $EW_{\rm \ion{Ca}{ii}}^{\rm IRT}$, 
$EW^{\rm H+K}_{\ion{Ca}{ii}}$, $EW_{\rm H\epsilon}$, and $V$ magnitude as a function of the rotational phase.} 
\label{fig:SAO51891_Halpha_He_CaII_Heps}
\end{figure}

The net equivalent width of these three lines (in particular the $\lambda$8498.06 line) shows a detectable 
rotational modulation, which becomes more evident if we consider the total emission of the triplet (see 
$EW_{\rm \ion{Ca}{ii}}^{\rm IRT}$ in the bottom panel in Fig.~\ref{fig:SAO51891_Halpha_He_CaII_Heps}). To evaluate 
the significance of the $EW_{\rm \ion{Ca}{ii}}^{\rm IRT}$ variation, we performed the $\chi^2$ 
test, as done for the $B-V$ index. The reduced $\chi^2$ of the fit is 0.19 for the Fourier polynomial fit and 
0.34 for the constant function. The probability that a fit of 5 free parameters gives $\chi^2<0.19$ is 0.08\%, 
while with one degree of freedom we obtain 44\%. This makes the rotational modulation of $EW_{\rm \ion{Ca}{ii}}^{\rm IRT}$ 
significant. The chromospheric radiative losses in the three lines of the \ion{Ca}{ii} IRT (Table~\ref{tab:fluxes}) have 
been calculated as for the H$\alpha$ and \ion{Ca}{ii} H\,\&\,K lines. They are nearly the same as those reported 
by \citet{Montes01a} and turn out to be only slightly larger than those we found for the H\,\&\,K lines. 

\subsection{Modeling the light and temperature curves}
\label{sec:spot_plage_model}
In \cite{Frasca05} we showed that, with an IDL spot model applied to contemporaneous 
light and temperature curves, it is possible to reconstruct the distribution of starspots and to remove the degeneracy 
of the spot parameters {\em temperature} and {\em area}. 
We modeled the observed light and temperature curves by assuming two active longitudes
on the stellar photosphere sketched by two circular regions whose 
flux contrast ($F_{\rm sp}/F_{\rm ph}$) can be evaluated through the Planck spectral energy distribution, the ATLAS9 
\citep{Kuru93} and PHOENIX NextGen \citep{Haus99} atmosphere models. 
Even if the real shape of the active regions can be very different, this approximation is useful to define
their main average parameters, like position, relative area, and temperature.
In \cite{Frasca05} we demonstrated that 
both the atmospheric models (ATLAS9 and NextGen) provide values of the spot temperature ($T_{\rm sp}$) and area 
coverage ($A_{\rm rel}$) in close agreement, while the black-body assumption for the SED leads to underestimate 
the spot temperature. Since we have no long-term record of the photospheric temperature, we assumed the maximum 
value obtained during our observations as the temperature of the unspotted photosphere. 

\begin{figure*}
\includegraphics[width=1.\textwidth]{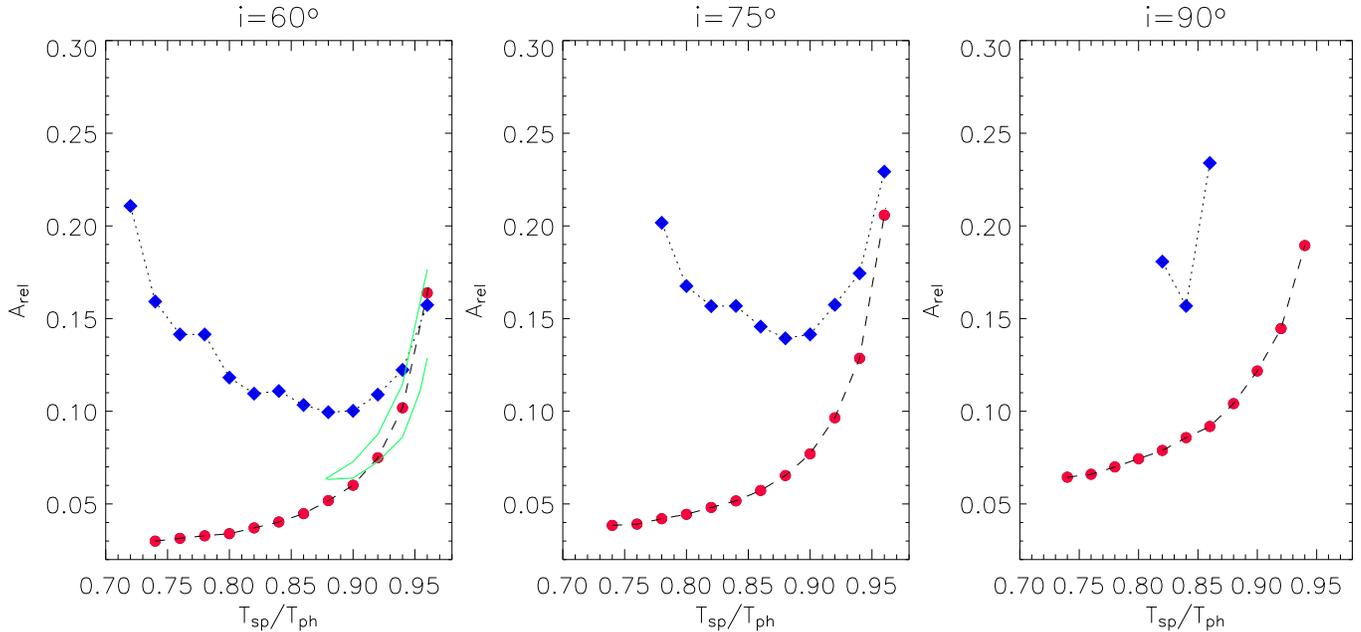}
\caption{Grids of solutions for SAO\,51891 obtained for three different values of $i$. The filled circles 
represent the values of spot temperature and area for the light-curve solutions, while the diamonds represent 
the solutions for the temperature-curve. The green lines in the left panel delimit the locus of the allowed 
solutions accounting for the data uncertainty.} 
\label{fig:solution_grids}
\end{figure*}

The application of our spot model requires the knowledge of the inclination of the rotation axis with respect to 
the line of sight. From our SED and the Tycho distance, we estimated a radius $R=0.54\pm0.09\,R_\odot$ 
(Table~\ref{tab:target_param}). By combining it with the rotation period of 2.42 days and the $v\sin i$ of 
19\,km\,s$^{-1}$ determined by us, we obtain a $\sin i>1$, testifying an inconsistency between these parameters.
We think that the small radius value, coming from the possibly underestimated Tycho distance, is the source of the
inconsistency, because the setback remains also with the lowest value of $v\sin i\simeq$\,15 km\,s$^{-1}$ from the 
literature.  If we use a typical radius for a K0V ZAMS star ($R=0.85\,R_\odot$; \citealt{Cox00}) we obtain an 
inclination $i$ of nearly $90{\degr}$, while adopting our highest radius estimate, $R=1.1\,R_\odot$, based on the 
distance of 50\,pc reported by \citet{Carpenter2008}, a value of $i\approx60{\degr}$ is found 
(Table~\ref{tab:literature_param}). Thus, we adopted three values of $i$ (namely $60{\degr}, 75{\degr}, 90{\degr}$) 
for our spot model. Then, from our unspotted temperature and magnitude, namely $T_{\rm ph}=5249$ K and $V=8\fm502$ 
we obtained, for each $i$, a grid of solutions for the temperature curve and another one for the light curve.
The limb-darkening coefficients $\mu_{6200}=0.55$ and $\mu_{\rm V}=0.74$ were used for the temperature
and the $V$ light curve, respectively (\citealt{AlNaimiy1978,Claret2000}). The intersection of the 
$T_{\rm eff}$ and $V$ grids provides us with the values of spot temperature and area. 
Figure \ref{fig:solution_grids} shows that, for any value of spot 
temperature ($T_{\rm sp}/T_{\rm ph}$), the size of the active regions must increase with the inclination in order to reproduce 
the observed variations. The effect is more pronounced for the temperature grid, so that we can have an intersection 
(a solution) only for $i=60{\degr}$, which occurs for $T_{\rm sp}/T_{\rm ph}=0.955$ and $A_{\rm rel}=0.149$. 
This seems to support a low inclination for the rotation axis, i.e. a large stellar radius. 
However, due to the strong uncertainty in the inclination, we are only able to give a very rough estimate 
of the spot temperature and area, while the longitudes of the two active regions (Table~\ref{tab:parameters_sao}), 
corresponding to 0.97 and 0.29 phases, are reliable.

We have also applied a plage model written in IDL (see, e.g., \citealt{Fra00}) to the modulation of the total 
\ion{Ca}{ii} IRT and HK emission with the aim of gaining some information about the surface inhomogeneities 
at chromospheric level. Given the scatter in the data, our code allowed us only to assess that the 
$EW^{\rm H+K}_{\ion{Ca}{ii}}$ and $EW^{\rm IRT}_{\ion{Ca}{ii}}$ curves are fairly reproducible with a single 
plage ($A_{\rm rel}\sim0.055$) placed between the two spots and closer to the smaller one. 
The only meaningful parameter that can be deduced from these data is the longitude of the chromospheric AR 
($\approx 110\degr$). However, this does not exclude two plages around the photospheric spots.

Our spot model allows us also to calculate the radial velocity variations induced by the line (and CCF) 
distortions owing to the passage of spots (or plages) over the star disk if the $v\sin i$ of the star is 
known (see \citealt{Frasca08b}). With the spot parameters listed in Table~\ref{tab:parameters_sao}, we 
calculated the expected RV curve and we superimposed it to the observed RV variations (with a full amplitude 
of about 500 m\,s$^{-1}$) in the bottom panel of Fig.~\ref{fig:palle_teff_V_err_sao}. 
The agreement with the data is apparent. This means that it is not necessary to invoke a low-mass companion 
(brown dwarf or giant planet in a close orbit) to explain the observed variations. Moreover, such variations 
prevent us to detect a Jovian planet around SAO\,51891, because the RV curve due to the planet with an amplitude of 
less than 100 m\,s$^{-1}$ would be lost inside the RV variations produced by the starspot.
Thus, very active stars, such as SAO\,51891, are not suitable targets for exo-planet search with the RV technique, 
unless a simultaneous detailed study of the magnetic activity phenomena is carried out.

\begin{table}  
\caption{Parameters of the two photospheric active regions.}
\label{tab:parameters_sao}
\begin{center}
\begin{tabular}{rcccc}
\hline
\hline
  \noalign{\smallskip}
   Lon.  &   Lat.  &  $\frac{T_{\rm sp}}{T_{\rm ph}}$ & $T_{\rm sp}$ &  $A_{\rm rel}$\\ 
  \noalign{\smallskip}
\hline
  \noalign{\smallskip}
   347$\degr$ & 62$\degr$ &  & &    \\
	       &	   & 0.955$_{-0.076}^{+0.045}$ & 5013$_{-350}^{+250}$ K & 0.149$_{-0.086}$ \\
   106$\degr$ & 39$\degr$ &  & &    \\
  \noalign{\smallskip}
\hline
\hline
\end{tabular}
\end{center}
\end{table}
\normalsize

\section{Conclusion}
\label{sec:concl}
In this paper we analyzed high-resolution {\it \'echelle} spectroscopic observations and contemporaneous $BV$ photometry of
the young star SAO\,51891. We report an updated spectral type, revised astrophysical parameters such as 
$T_{\rm eff}$, $\log g$, [Fe/H], rotational and heliocentric radial velocity, space motion, 
spectral energy distribution, and lithium abundance. Our main goal was to investigate short-term 
variability due to magnetic activity at both photospheric and chromospheric levels. 

The kinematics, the lithium abundance, and the level of photospheric and chromospheric activity 
indicate an age similar to that of the Pleiades cluster. We also confirm the membership in the Local Association,
with an age of the order of 100 Myr. The spectral energy distribution provides evidence for lacking of 
significant amounts of circumstellar dust and upper limits are derived. 
The small-amplitude ($\approx$\,500 m\,s$^{-1}$) radial velocity variations measured from our 
spectra are correlated with the stellar rotation and fully explained with the line distortions produced 
by the starspots. Thus there is no evidence for a stellar or sub-stellar companion, based on radial velocity 
measurements or direct imaging. Our analysis demonstrates the difficulty of detecting sub-stellar companions 
(brown dwarfs or giant planets) around very active stars from radial velocity measurements.

From our spectra, covering one and a half stellar rotation, we find conspicuous chromospheric emission 
in the \ion{Ca}{ii} H\,\&\,K, \ion{Ca}{ii} IRT  and H$\epsilon$ lines. The cores of the H$\alpha$ line 
are also clearly filled in by chromospheric emission. All these 
features confirm the strong magnetic activity of SAO\,51891.

We detected a clear modulation of the $V$ and $T_{\rm eff}$ curves due to spots, and a low-amplitude modulation 
of the \ion{Ca}{ii} IRT, H\,\&\,K, and H$\epsilon$ emissions ascribed to chromospheric inhomogeneities. Astonishingly, 
we did not find any clear modulation in H$\alpha$, in contrast to what we observed for several other active stars.
We speculate that the modulation produced by plages is possibly hidden by other activity signatures, like 
microflares, which can significantly affect the H$\alpha$ emission.

The simultaneous solution of the $V$ light-curve and the temperature modulation with our spot model allowed us 
to reconstruct the approximate spot distribution on the stellar photosphere by adopting an inclination 
$i=60{\degr}$. We found two fairly large active longitudes with a temperature close to the ``unspotted'' effective temperature 
($\Delta T\simeq$\,240\,K), different to what we have found for other mildly-active main sequence stars.
We want to stress that our model gives us only a ``rough'' estimation of the inhomogeneities parameters mainly for the
following reasons: i) there is a phase gap in our data (in particular the spectroscopic observations) just after the maximum 
of the curves which can introduce some uncertainty on the ``unspotted'' level of temperature; ii) uncertainties on the 
distance translate into uncertainties in the stellar radius and, consequently, the inclination angle cannot be constrained 
with sufficient accuracy.

To date, photometric and spectroscopic analysis comparable to that reported here has been conducted on only a handful of 
young stars. In order to place our own solar system in context, understand the range of planet formation outcomes as a 
function of stellar parameters, as well as investigate the early angular momentum evolution of sun-like stars, more observations 
of this kind are required. For this reason, we already started a program of high-resolution FOCES@CAHA and SARG@TNG {\it \'echelle} 
spectroscopic observations of young late-type stars to improve our knowledge of photospheric/chromospheric 
inhomogeneities of this stellar population.

\begin{acknowledgements}

The authors are very grateful to the referee Ilya Yu Alekseev for a careful reading of the paper and valuable comments. 
We thank the Calar Alto Observatory and OACt teams for the assistance during the observations. This work has been supported by the 
Italian {\em Ministero dell'Istruzione, Universit\`a e  Ricerca} (MIUR) which is gratefully acknowledged. KB has been 
also supported by the ESO DGDF 2008. EC \& JMA acknowledge financial support from INAF (PRIN 2007: From active accretion to debris 
discs). This research was based on SIMBAD and VIZIER databases, operated at CDS (Strasbourg, France), and on INES (IUE New Extracted 
Spectra) data from the IUE satellite.

\end{acknowledgements}

\appendix{
\section{Spectra of SAO\,51891 in the H$\alpha$ and \ion{Ca}{ii} IRT regions, and calcium EWs values.}
\label{appendix:a}

\begin{figure}[h]
\centering
\includegraphics[width=0.5\textwidth]{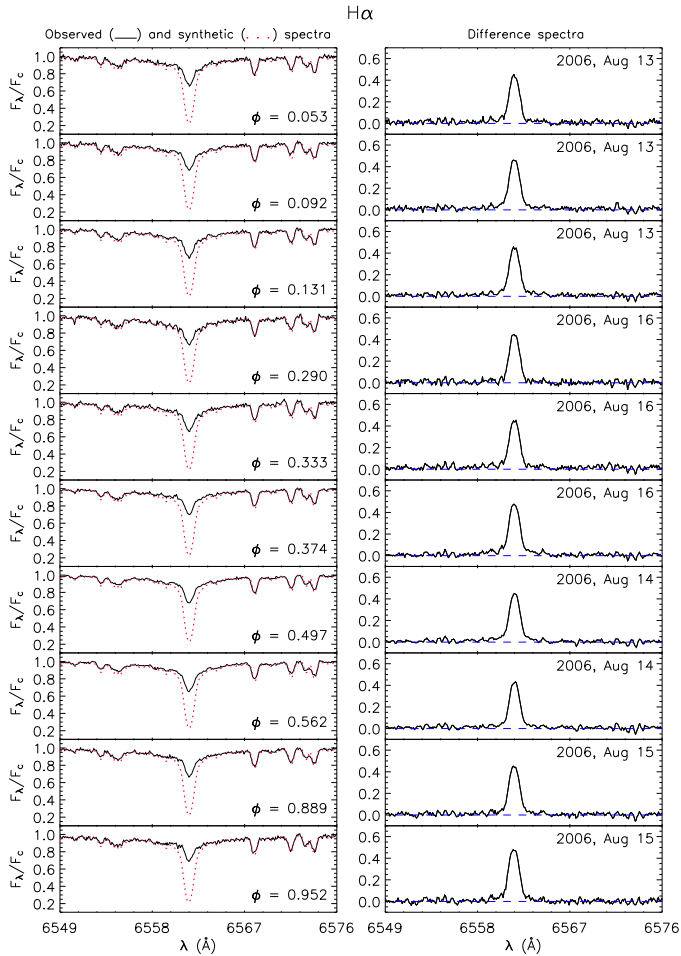}
\caption{{\it Left panels}: observed and continuum-normalized spectra of SAO\,51891 in the H$\alpha$ region together 
with the non-active template. {\it Right panels}: difference between observed and template spectra.} 
\label{fig:SAO51891_spettri2006_Halpha}
\end{figure}

\begin{figure}[h]
\centering
\includegraphics[width=0.5\textwidth]{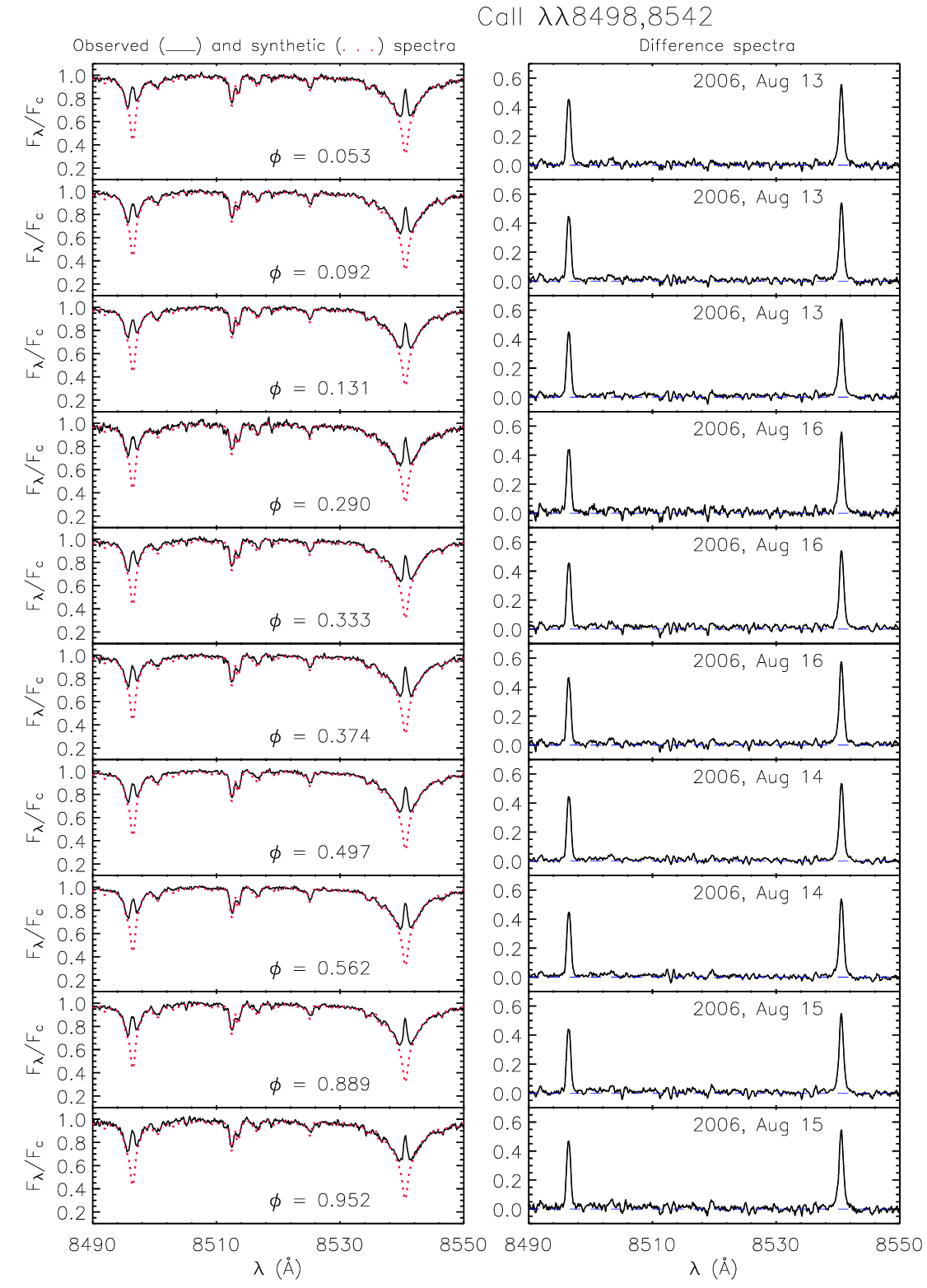}
\caption{{\it Left panels}: observed and continuum-normalized spectra of SAO\,51891 in the \ion{Ca}{ii} IRT 
$\lambda\lambda8498,8542$ region together with the non-active template. {\it Right panels}: difference between observed and template 
spectra.} 
\label{fig:SAO51891_spettri2006_CaII8500}
\end{figure}

\begin{figure}[h]
\centering
\includegraphics[width=0.5\textwidth]{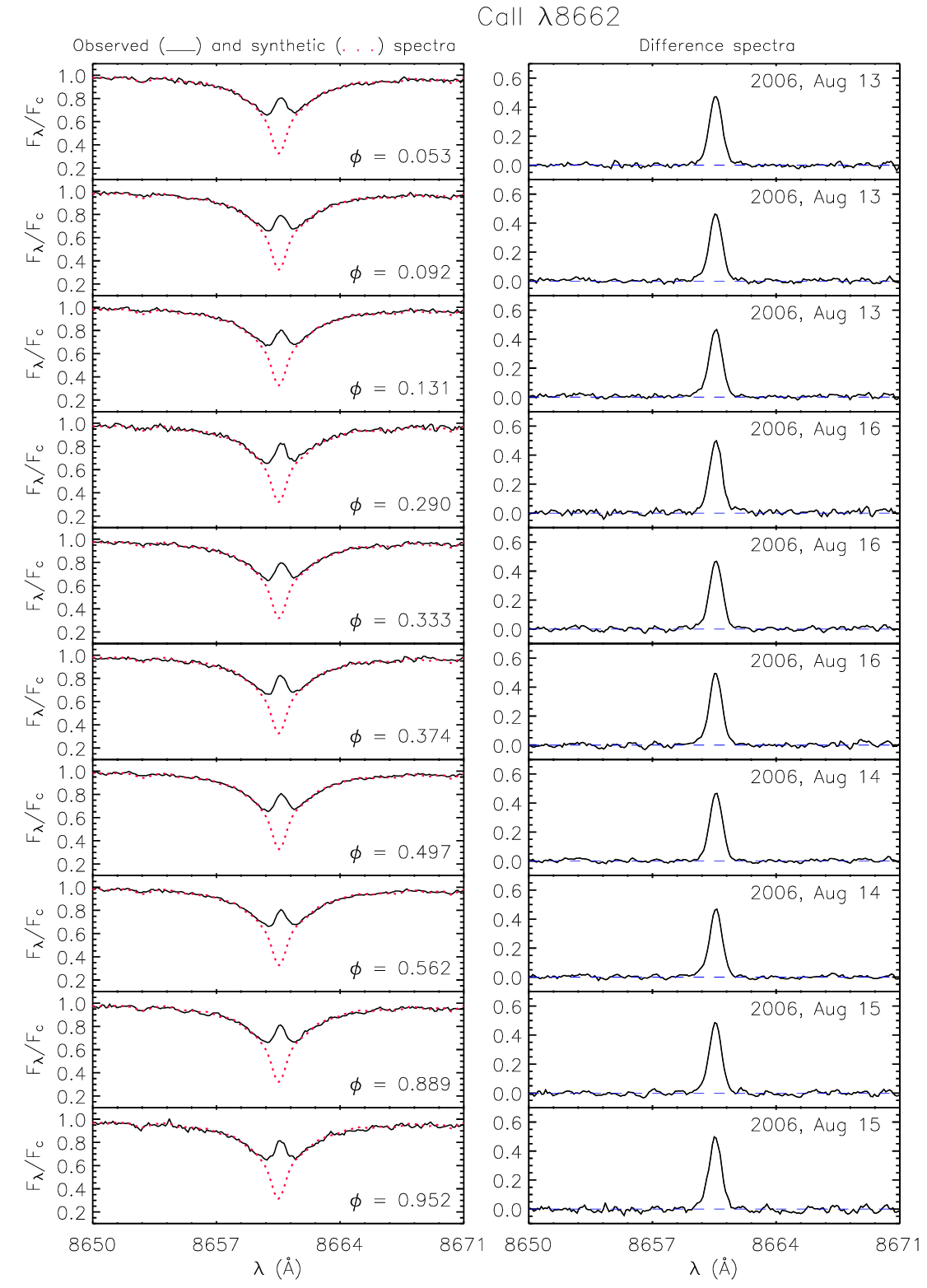}
\caption{{\it Left panels}: observed and continuum-normalized spectra of SAO\,51891 in the \ion{Ca}{ii} IRT $\lambda8662$ 
region together with the non-active template. {\it Right panels}: difference between observed and template spectra.} 
\label{fig:SAO51891_spettri2006_CaII8662}
\end{figure}

\begin{table*}[t]   
\caption{$EW$ values of  the \ion{Ca}{ii} lines.}
\label{tab:calcium_line}
\scriptsize
\begin{center}
\begin{tabular}{ccccccc}
\hline
\hline
  \noalign{\smallskip}
HJD & Phase & $EW_{\rm\ion{Ca}{ii}}^{\lambda 8498}$ & $EW_{\rm\ion{Ca}{ii}}^{\lambda 8542}$ &$EW_{\rm\ion{Ca}{ii}}^{\lambda 8662}$&$EW_{\rm \ion{Ca}{ii}}^{\rm
K}$&$EW_{\rm \ion{Ca}{ii}}^{\rm H}$\\ 
(+2\,400\,000) & & (\AA) & (\AA) & (\AA) & (\AA) & (\AA)\\ 
  \noalign{\smallskip}
\hline
  \noalign{\smallskip}
53\,961.379 & 0.053 & 0.406$\pm$0.025 & 0.554$\pm$0.040 & 0.464$\pm$0.029 & 0.845$\pm$0.383 & 0.665$\pm$0.130 \\
53\,961.473 & 0.092 & 0.422$\pm$0.033 & 0.536$\pm$0.034 & 0.437$\pm$0.024 & 0.677$\pm$0.048 & 0.491$\pm$0.035 \\
53\,961.566 & 0.131 & 0.406$\pm$0.028 & 0.545$\pm$0.034 & 0.455$\pm$0.027 & 0.654$\pm$0.030 & 0.541$\pm$0.046 \\
53\,962.453 & 0.497 & 0.412$\pm$0.031 & 0.558$\pm$0.037 & 0.434$\pm$0.025 & 0.578$\pm$0.034 & 0.538$\pm$0.031 \\
53\,962.609 & 0.562 & 0.419$\pm$0.034 & 0.551$\pm$0.039 & 0.439$\pm$0.029 & 0.564$\pm$0.059 & 0.471$\pm$0.034 \\
53\,963.402 & 0.889 & 0.396$\pm$0.028 & 0.530$\pm$0.037 & 0.434$\pm$0.025 & 0.607$\pm$0.044 & 0.500$\pm$0.029 \\
53\,963.555 & 0.952 & 0.389$\pm$0.032 & 0.510$\pm$0.047 & 0.438$\pm$0.035 & 0.684$\pm$0.047 & 0.547$\pm$0.050 \\
53\,964.371 & 0.290 & 0.416$\pm$0.046 & 0.546$\pm$0.041 & 0.469$\pm$0.034 & 0.703$\pm$0.100 & 0.587$\pm$0.064 \\
53\,964.477 & 0.333 & 0.429$\pm$0.036 & 0.547$\pm$0.037 & 0.444$\pm$0.032 & 0.699$\pm$0.054 & 0.510$\pm$0.027 \\
53\,964.574 & 0.374 & 0.422$\pm$0.036 & 0.579$\pm$0.038 & 0.462$\pm$0.030 & 0.676$\pm$0.041 & 0.554$\pm$0.033 \\
  \noalign{\smallskip}
\hline
\hline
\end{tabular}
\end{center}
\end{table*}
\normalsize

\section{Photometric data}
\label{appendix:b}

\begin{table*}[t]   
\caption{$V$ and $B-V$ values.}
\label{tab:phot_data}
\scriptsize
\begin{center}
\begin{tabular}{cccc}
\hline
\hline
  \noalign{\smallskip}
HJD            & Phase & $V$   & $B-V$ \\ 
(+2\,400\,000) &       & (mag) & \\ 
  \noalign{\smallskip}
\hline
  \noalign{\smallskip}
53\,962.50219 & 0.517 & 8.519$\pm$0.005 & 0.803$\pm$0.010 \\
53\,962.55728 & 0.540 & 8.527$\pm$0.007 & 0.800$\pm$0.009 \\
53\,962.58972 & 0.554 & 8.523$\pm$0.006 & 0.794$\pm$0.005 \\
53\,963.56109 & 0.955 & 8.564$\pm$0.005 & 0.811$\pm$0.004 \\
53\,963.57771 & 0.962 & 8.568$\pm$0.009 & 0.807$\pm$0.011 \\
53\,963.58683 & 0.966 & 8.571$\pm$0.005 & 0.807$\pm$0.009 \\
53\,963.59494 & 0.969 & 8.560$\pm$0.006 & 0.812$\pm$0.009 \\
53\,964.55451 & 0.366 & 8.548$\pm$0.005 & 0.806$\pm$0.009 \\
53\,964.56212 & 0.369 & 8.539$\pm$0.013 & 0.810$\pm$0.014 \\
53\,964.56986 & 0.372 & 8.539$\pm$0.011 & 0.808$\pm$0.008 \\
53\,964.57734 & 0.375 & 8.540$\pm$0.010 & 0.802$\pm$0.004 \\
53\,964.58673 & 0.379 & 8.553$\pm$0.001 & 0.810$\pm$0.010 \\
53\,964.59357 & 0.382 & 8.537$\pm$0.010 & 0.802$\pm$0.006 \\
53\,965.55485 & 0.779 & 8.538$\pm$0.010 & 0.794$\pm$0.008 \\
53\,965.56308 & 0.782 & 8.533$\pm$0.005 & 0.804$\pm$0.006 \\
53\,965.57115 & 0.786 & 8.535$\pm$0.009 & 0.802$\pm$0.005 \\
53\,965.57839 & 0.789 & 8.538$\pm$0.006 & 0.802$\pm$0.007 \\
53\,965.60166 & 0.798 & 8.537$\pm$0.003 & 0.804$\pm$0.006 \\
53\,966.55439 & 0.192 & 8.550$\pm$0.007 & 0.805$\pm$0.009 \\
53\,966.57410 & 0.200 & 8.576$\pm$0.007 & 0.807$\pm$0.003 \\
53\,966.58242 & 0.203 & 8.563$\pm$0.008 & 0.815$\pm$0.007 \\
53\,966.58971 & 0.206 & 8.572$\pm$0.005 & 0.807$\pm$0.011 \\
53\,967.55227 & 0.604 & 8.502$\pm$0.009 & 0.802$\pm$0.004 \\
53\,969.54477 & 0.428 & 8.530$\pm$0.004 & 0.803$\pm$0.004 \\
53\,969.55438 & 0.432 & 8.528$\pm$0.004 & 0.803$\pm$0.012 \\
53\,969.56224 & 0.435 & 8.532$\pm$0.004 & 0.805$\pm$0.012 \\
53\,969.57359 & 0.440 & 8.534$\pm$0.005 & 0.800$\pm$0.009 \\
53\,969.58526 & 0.444 & 8.546$\pm$0.007 & 0.790$\pm$0.008 \\
  \noalign{\smallskip}
\hline                                   
\hline                                   
\end{tabular}
\end{center}
\end{table*}
\normalsize
}

\bibliographystyle{aa}

\end{document}